\documentclass[showpacs,preprintnumbers,amsmath,amssymb,superscriptaddress]{revtex4}

\usepackage{graphicx}
\usepackage{dcolumn}
\usepackage{bm}
\usepackage{epsfig}
\usepackage{subfigure}

\newcommand\independent{\protect\mathpalette{\protect\independenT}{\perp}}
\def\independenT#1#2{\mathrel{\setbox0\hbox{$#1#2$}%
\copy0\kern-\wd0\mkern4mu\box0}}

\begin{document}

\title{Persistence of nonlocality for bipartite quantum spin systems}

\author{Kaushik Parasuram S.}\email{kaushik.parasuram@gmail.com}
\affiliation{Birla Institute of Technology $\&$ Science-Pilani, Pilani - 333 031, Rajasthan, India.}
\affiliation{The Institute of Mathematical Sciences, C. I. T. Campus, Taramani, Chennai - 600 113, India.}

\author{Sibasish Ghosh} \email{sibasish@imsc.res.in}
\affiliation{The Institute of Mathematical Sciences, C. I. T. Campus, Taramani, Chennai - 600 113, India.}

\begin{abstract}
Generic forms of the entangled states of two spin-$1$ (and spin-$\frac{3}{2}$) particles, along with the set of appropriate spin observables that together exhibit maximum nonlocality under the Hardy's nonlocality test are given; the maximum nonlocality is shown to be $0.09017$. It is conjectured that this result holds good for a system of two spin-$j$ particles for all values of $j$. It is also shown that no maximally entangled state of two spin-$j$ particles responds to the Hardy's nonlocality test.
\end{abstract}


\maketitle

\section{Introduction}
The Hardy's nonlocality test is an experiment that showcases the nonlocality present in entangled states of bipartite as well as multipartite quantum systems. Unlike the Bell test, it is devoid of inequalities. Hardy originally gave the test for a  system comprising of the spins of an electron and positron in an experimental setup consisting of two overlapping Mach-Zehnder interferometers \cite{hardy1}. Clifton and Niemann \cite{cliff1} generalized it for a system of two spin-$j$ particles, which was later reduced to it's minimal form by Kunkri and Choudhary \cite{kunkri}. This later version of the Hardy's nonlocality test for two spin-$j$ particles prescribes a minimal number of non-commuting spin-$j$ observables (two for each spin-$j$ particle) and a minimal number of correlation measurements on the two particles such that one of the correlations gives a measure of the amount of nonlocality. The conditions on the correlation measurements are such that the four spin-$j$ observables, if replaced by $(2j+1)$-valued classical random variables, yield an incompatible set of joint probabilities, thus revealing the nonclassical nature of the quantum correlation present in the joint state of the two spin-$j$ particles.

It is known that the Hardy's nonlocality test shows a maximum nonlocality (in terms of a joint probability) of $(- 11 + 5{\sqrt{5}})/2 \approx 0.09017$, over all possible states of two spin-$\frac{1}{2}$ particles, for all possible choices of observables; and that the maximally entangled states of the system do not respond to the test. In this paper, we explicitly work out the test for systems consisting of two spin-$1$ and spin-$\frac{3}{2}$ particles; we obtain their entangled states along with the sets of appropriate observables that together exhibit maximum nonlocality. The maximum nonlocality in these systems turns out to be $(- 11 + 5{\sqrt{5}})/2$ as well and we conjecture that this result holds good for all higher spin values. Validity of this conjecture shows ``persistence of nonlocality'' as opposed to the general notion that higher the spin value $j$, more closer will be the behaviour of the spin-$j$ quantum system to a classical one \cite{cliff1}. We also show that no maximally entangled state of two spin-$j$ particles responds to the Hardy's nonlocality test.

In section ${\rm \ref{test}}$, the Hardy's nonlocality test for two $(2j+1)$-level systems is described. In section ${\rm \ref{maxent}}$, it is shown that no maximally entangled state of two spin-$j$ particles satisfies the Hardy's nonlocality conditions. In sections ${\rm \ref{spin1by2}}$, ${\rm \ref{spin1}}$ and ${\rm \ref{spin3by2}}$, the Hardy's nonlocality test for systems of two spin-${\rm \frac{1}{2}}$, spin-$1$ and spin-$\frac{3}{2}$ particles are described, respectively. Having found the maximally nonlocal Hardy state with the same amount of maximum nonlocality for $j = \frac{1}{2}$, $1$, $\frac{3}{2}$, in section ${\rm \ref{general}}$, the generic form of such a Hardy state for {\it any} spin $j$ and it's maximum amount of nonlocality is discussed and a conjecture made on the persistence of nonlocality. The result is verified with a discussion of the test for a system of two spin-$2$ particles in Appendix D. Section ${\rm \ref{concl}}$ deals with the conclusion. Some proofs are given in a few other appendices.

\section{Hardy's nonlocality test for two $(2j+1)$-level systems}
\label{test}
The Hardy's nonlocality test for two $(2j+1)$-level systems, in it's minimal form, is given by the following conditions:
\begin{equation}
\label{hardygen}
\left.
\begin{array}{lcl}
{\rm prob}(A_1=+j,\ B_1=+j) & = & 0,\\
{\rm prob}(A_1=+j-1,\ B_2=-j) & = & 0,\\
{\rm prob}(A_1=+j-2,\ B_2=-j) & = & 0,\\
\ldots & \ldots & \ldots \\
\ldots & \ldots & \ldots \\
{\rm prob}(A_1=-j,\ B_2=-j) & = & 0,\\
{\rm prob}(A_2=-j,\ B_1=+j-1) & = & 0,\\
{\rm prob}(A_2=-j,\ B_1=+j-2) & = & 0,\\
\ldots & \ldots & \ldots \\
\ldots & \ldots & \ldots \\
{\rm prob}(A_2=-j,\ B_1=-j) & = & 0,\\
{\rm prob}(A_2=-j,\ B_2=-j) & \equiv & q > 0,
\end{array}
\right \}
\end{equation}
where each of $A_1$, $A_2$, $B_1$ and $B_2$ is a $\{+j,\ +j-1,\ ...,\ -j\}$-valued random variable.

No local realistic theory satisfies these conditions \cite{kunkri}. However, there exist quantum mechanical states of two spin-$j$ particles for which these conditions are satisfied when the random variables $A_1$, $A_2$, $B_1$ and $B_2$ are replaced by spin-$j$ observables $\hat{A}_1$, $\hat{A}_2$, $\hat{B}_1$ and $\hat{B}_2$, respectively, such that $\hat{A}_1\&\hat{A}_2$ (and $\hat{B}_1\&\hat{B}_2$) are non-commuting. Evidently, the $(4j + 2)$-th state $|\hat{A}_2 = - j,\ \hat{B}_2 = - j\rangle$ has to be linearly independent of the first $(4j + 1)$ number of states $|\hat{A}_1 =  j,\ \hat{B}_1 =  j\rangle$, $|\hat{A}_1 =  j - 1,\ \hat{B}_2 = - j\rangle$, $|\hat{A}_1 =  j - 2,\ \hat{B}_2 = - j\rangle$, $\ldots$, $|\hat{A}_1 =  - j,\ \hat{B}_2 = - j\rangle$,    $|\hat{A}_2 = - j,\ \hat{B}_1 = j - 1\rangle$, $|\hat{A}_2 =  - j,\ \hat{B}_1 = j - 2\rangle$, $\ldots$, $|\hat{A}_2 =  - j,\ \hat{B}_1 = - j\rangle$ in order that $q > 0$. So, for a given set of two pairs of non-commuting spin-$j$ observables $(\hat{A}_1,\ \hat{A}_2)$ and $(\hat{B}_1,\ \hat{B}_2)$, the maximum value of $q$ will be attained by that state, which is obtained from the $(4j + 2)$-th state by projecting out the contributions from the first $(4j + 1)$ states in it. This approach will not help us to test whether an arbitrarily given entangled state of two spin-$j$ systems would respond positively to the Hardy's nonlocality test given by Eq.(\ref{hardygen}). Keeping this fact in mind, apart from the maximum value of $q$, we provide in this paper (for a given set of observables), the set of {\it all} states of two spin-$j$ particles, each of which will satisfy the conditions in Eq.(\ref{hardygen}), for a few smaller values of $q$.

In this direction, consider Alice and Bob, who are two far-apart observers, sharing multiple identical copies of an entangled state of two spin-$j$ particles between them. Say Alice can perform measurement of any one of the two non-commuting spin-$j$ observables $\hat{A}_1\&\hat{A}_2$ on her spin-$j$ particle, while Bob can perform measurement of any one of the two non-commuting observables $\hat{B}_1\&\hat{B}_2$ on his spin-$j$ particle. Without any loss of generality, we let Alice and Bob fix one of their observables along their own independent (and arbitrary) z-axes i.e. $\hat{A}_2 \equiv \hat{S}_z$ and $\hat{B}_2 \equiv \hat{S}_z$. The observables $\hat{A}_1$ and $\hat{B}_1$ are then chosen to be of the form:
\begin{eqnarray*}
&&\hat{A}_1=\hat{m}.\hat{S},\\
&&\hat{B_1}= \hat{n}.\hat{S},
\end{eqnarray*}
where $\hat{m}=(\sin\theta_1\cos\phi_1,\ \sin\theta_1\sin\phi_1,\ \cos\theta_1),\ \hat{n}=(\sin\theta_2\cos\phi_2,\ \sin\theta_2\sin\phi_2,\ \cos\theta_2)$, with $\theta_1,\ \theta_2 \ \in (0,\ \pi)$ and $\phi_1,\ \phi_2 \ \in [0,\ 2\pi)$; $\hat{S}=(\hat{S}_x,\ \hat{S}_y,\ \hat{S}_z)$. Note that the two end points $\theta_1=0,\ \pi$ (and also $\theta_2=0,\ \pi$) are not considered here as $\hat{A}_1\&\hat{A}_2$ (and also $\hat{B}_1\&\hat{B}_2$) are taken to be non-commuting. The two observers then perform local measurements on each copy of the shared entangled state. We call these measurements here as {\it correlation-measurements}. 

\section{Hardy's nonlocality for maximally entangled states in ${C\!\!\!\!I}^{2j+1} \otimes {C\!\!\!\!I}^{2j+1}$}
\label{maxent}
Here, we show that there exists no maximally entangled state of two spin-$j$ particles that responds to the Hardy's nonlocality test.

\textbf{Proof:}
Assume $|\Psi\rangle$ to be a maximally entangled states in ${C\!\!\!\!I}^{2j+1} \otimes {C\!\!\!\!I}^{2j+1}$. Let
\begin{eqnarray*}
&|\Psi_0\rangle&=\frac{1}{\sqrt{2j+1}}\displaystyle\sum_{s \in \{-j,\ -j+1,.. ,\ +j\}}|\hat{A_2}=s\rangle\otimes|\hat{B_2}=s\rangle
\end{eqnarray*}
be the standard maximally entangled state in ${C\!\!\!\!I}^{2j+1} \otimes {C\!\!\!\!I}^{2j+1}$. So, any general maximally entangled state $|\Psi\rangle$ can then be written as $|\Psi\rangle = (I\otimes U)|\Psi_0\rangle,$ where $U$ is a $(2j+1)\times (2j+1)$ unitary matrix. Thus, we have here
\begin{eqnarray}
&|\Psi\rangle&=\frac{1}{\sqrt{2j+1}}\displaystyle\sum_{s \in \{-j,\ -j+1,.. ,\ +j\}}|\hat{A_2}=s\rangle\otimes U|\hat{B_2}=s\rangle.
\label{maxentangled}
\end{eqnarray}

The Hardy's nonlocality test in Eq.(\ref{hardygen}), when applied to $|\Psi\rangle$, gives rise to the following conditions:
\begin{equation}
\label{orthoj1}
|\langle\Psi|\hat{A_1}=+j,\ \hat{B_1}=+j\rangle|^2 = 0,
\end{equation}
\begin{equation}
\label{orthoj2}
\left.
\begin{array}{lcl}
|\langle\Psi|\hat{A_1}=+j-1,\ \hat{B_2}=-j\rangle|^2 & = & 0,\\
|\langle\Psi|\hat{A_1}=+j-2,\ \hat{B_2}=-j\rangle|^2 & = & 0,\\
\ldots & \ldots & \ldots\\
\ldots & \ldots & \ldots \\
|\langle\Psi|\hat{A_1}=-j,\ \hat{B_2}=-j\rangle|^2 & = & 0,
\end{array}
\right \}
\end{equation}
\begin{equation}
\label{orthoj3}
\left.
\begin{array}{lcl}
|\langle\Psi|\hat{A_2}=-j,\ \hat{B_1}=+j-1\rangle|^2 & = & 0,\\
|\langle\Psi|\hat{A_2}=-j,\ \hat{B_1}=+j-2\rangle|^2 & = & 0,\\
\ldots & \ldots & \ldots \\
\ldots & \ldots & \ldots \\
|\langle\Psi|\hat{A_2}=-j,\ \hat{B_1}=-j\rangle|^2 & = & 0,
\end{array}
\right \}
\end{equation}
\begin{equation}
\label{orthoj4}
|\langle\Psi|\hat{A_2}=-j,\ \hat{B_2}=-j\rangle|^2 \equiv q >0.
\end{equation}

From the conditions in Eq.(\ref{orthoj3}), we get:
\begin{equation*}
\left.
\begin{array}{lcl}
|\langle \hat{B_1}=+j-1|U|\hat{B_2}=-j\rangle|^2 & = & 0,\\
|\langle \hat{B_1}=+j-2|U|\hat{B_2}=-j\rangle|^2 & = & 0,\\
\ldots & \ldots & \ldots \\
\ldots & \ldots & \ldots \\
|\langle \hat{B_1}=-j|U|\hat{B_2}=-j\rangle|^2 & = & 0.
\end{array}
\right \}
\end{equation*}

These conditions imply that
\begin{equation}
\label{maxent1}
U|\hat{B_2}=-j\rangle \propto |\hat{B_1}=+j\rangle.
\end{equation}

From the condition in Eq.(\ref{orthoj1}), we get:
\begin{eqnarray}
&&|\frac{1}{\sqrt{2j+1}}\displaystyle\sum_{s \in \{-j,\ -j+1,.. ,\ +j\}}\langle\hat{A_2}=s|\hat{A_1}=+j\rangle\times\langle\hat{B_2}=s|U^\dagger|\hat{B_1}=+j\rangle|^2=0.
\label{maxent2}
\end{eqnarray}

From Eq.(\ref{maxent1}) and Eq.(\ref{maxent2}), we get:
\begin{eqnarray}
|\langle\hat{A_2}=-j|\hat{A_1}=+j\rangle|^2=0.
\label{maxent3}
\end{eqnarray}

From Eq.(\ref{orthoj2}), for all $s \in \{-j,\ -j+1,.. ,\ +j-1\}$, we get:
\begin{eqnarray}
&&|\langle\Psi|\hat{A_1}=s,\ \hat{B_2}=-j\rangle|^2=0,\nonumber\\
\Rightarrow && |\frac{1}{\sqrt{2j+1}}\displaystyle\sum_{t \in \{-j,\ -j+1,.. ,\ +j\}}\langle\hat{A_2}=t|\hat{A_1}=s\rangle\times\langle\hat{B_2}=t|U^\dagger|\hat{B_2}=-j\rangle|^2=0.
\label{maxent4}
\end{eqnarray}

Thus, for all $s \in \{-j,\ -j+1,.. ,\ +j-1\}$, we have:
\begin{eqnarray}
|(\displaystyle\sum_{t \in \{-j,\ -j+1,.. ,\ +j\}}\langle\hat{B_2}=-j|U|\hat{B_2}=t\rangle|\hat{A_2}=t\rangle)^\dagger|\hat{A_1}=s\rangle|&=0,\nonumber\\
\Rightarrow\ |\hat{A_1}=s\rangle \perp \displaystyle\sum_{t \in \{-j,\ -j+1,.. ,\ +j\}}\langle\hat{B_2}=-j|U|\hat{B_2}=t\rangle|\hat{A_2}=t\rangle,\nonumber\\
\Rightarrow\ \displaystyle\sum_{t \in \{-j,\ -j+1,.. ,\ +j\}}\langle\hat{B_2}=-j|U|\hat{B_2}=t\rangle|\hat{A_2}=t\rangle \propto |\hat{A_1}=+j\rangle.
\label{maxent5}
\end{eqnarray}

Using Eq.(\ref{maxent3}) in Eq.(\ref{maxent5}), we get:
\begin{eqnarray}
&&\langle\hat{B_2}=-j|U|\hat{B_2}=-j\rangle=0.
\label{maxent6}
\end{eqnarray}

From Eq.(\ref{orthoj4}) and Eq.(\ref{maxent6}), we get:
\begin{eqnarray}
0<q&=&|\frac{1}{\sqrt{2j+1}}\displaystyle\sum_{t \in \{-j,\ -j+1,.. ,\ +j\}}\langle\hat{A_2}=t|\hat{A_2}=-j\rangle\times\langle\hat{B_2}=t|U^\dagger|\hat{B_2}=-j\rangle|^2\nonumber\\
&=&\frac{1}{\sqrt{2j+1}}|\langle\hat{B_2}=-j|U^\dagger|\hat{B_2}=-j\rangle|^2\nonumber\\
&=&\frac{1}{\sqrt{2j+1}}|\overline{\langle\hat{B_2}=-j|U|\hat{B_2}=-j\rangle}|^2\nonumber\\
&=&0.
\label{maxent7}
\end{eqnarray}

So, $q$ must be zero--a contradiction. Hence, it is proved that there exists no maximally entangled state of two spin-$j$ particles that satisfies the Hardy's nonlocality conditions. \textbf{QED}

\section{A system of two spin-$\frac{1}{2}$ particles}
\label{spin1by2}
For the sake of continuity of argument, in this section we briefly mention the Hardy's nonlocality test for states of two spin-$\frac{1}{2}$ particles, which is already known in the literature. Note that in all calculations pertaining to this system, the eigenvalues of the spin-$\frac{1}{2}$ observables are taken to be $\pm 1$ instead of $\pm\frac{1}{2}$.

The spin-$\frac{1}{2}$ observables $\hat{A}_1$ and $\hat{B}_1$ may be written in their most general form, in the eigen-bases of $\hat{A}_2 \equiv \hat{S_z}$ and $\hat{B}_2 \equiv \hat{S_z}$, as:
\[\hat{A}_1=\left( \begin{array}{cc}
\cos\theta_1 & \sin\theta_1 e^{-i \phi_1} \\
\sin\theta_1 e^{i \phi_1} & -\cos\theta_1 \end{array} \right),\
\hat{B}_1=\left( \begin{array}{cc}
\cos\theta_2 & \sin\theta_2 e^{-i \phi_2} \\
\sin\theta_2 e^{i \phi_2} & -\cos\theta_2 \end{array} \right).\]

Their eigenvectors are then given by:
\begin{equation}
\label{eigenvec1by2}
\left.
\begin{array}{lcl}
|\hat{A}_1=+1\rangle=a_{11}|+1\rangle +a_{12}|-1\rangle,\\
|\hat{A}_1=-1\rangle=a_{21}|+1\rangle +a_{22}|-1\rangle,\\
|\hat{B}_1=+1\rangle=b_{11}|+1\rangle +b_{12}|-1\rangle,\\
|\hat{B}_1=-1\rangle=b_{21}|+1\rangle +b_{22}|-1\rangle,
\end{array}
\right \}
\end{equation}
where
\[\left( \begin{array}{cc}
a_{11} & a_{12}\\
a_{21} & a_{22} \end{array} \right) =\left( \begin{array}{cc}
\cos \frac{\theta_1}{2}& e^{i \phi_1} \sin \frac{\theta_1}{2}\\
e^{-i \phi_1} \sin \frac{\theta_1}{2}& -\cos \frac{\theta_1}{2} \end{array} \right)\]
and similar expressions for $b_{ij}$'s with $({\theta}_1,\ {\phi}_1)$ being replaced by $({\theta}_2,\ {\phi}_2)$. In this system, non-commutativity of the observables at each observer demands that the eigenvectors of each observable have a strictly non-zero projection on the eigenvectors of the other observable at the same observer i.e. if $|\hat{A}_i=\pm{1}\rangle=\alpha^{\pm}|\hat{A}_j=+1\rangle+\beta^{\pm}|\hat{A}_j=-1\rangle$ with $i\not=j$, then $\alpha^{\pm}\beta^{\pm}\not=0$. Hence, all $a_{ij}$'s and $b_{ij}$'s ($i,\ j \ \in \{1,\ 2\}$) in Eq.(\ref{eigenvec1by2}) are strictly non-zero.

\subsection{The Hardy's nonlocality test}
The following conditions constitute the Hardy's nonlocality test for the given system:
\begin{equation}
\label{prob1by2}
\left.
\begin{array}{lcl}
\rm{prob}(\hat{A}_1=+1,\ \hat{B_1}=+1)=0, \\
\rm{prob}(\hat{A}_1=-1,\ \hat{B_2}=-1)=0, \\
\rm{prob}(\hat{A}_2=-1,\ \hat{B_1}=-1)=0, \\
\rm{prob}(\hat{A}_2=-1,\ \hat{B_2}=-1)=q > 0.
\end{array}
\right \}
\end{equation}
One can show that the four conditions in Eq.(\ref{prob1by2}) will have contradiction with local-realism.

We associate a state with every condition in the test, say:
\begin{equation}
\label{ket1by2}
\left.
\begin{array}{lcl}
|\Phi_1\rangle=|\hat{A}_1=+1\rangle\otimes|\hat{B}_1=+1\rangle,\\
|\Phi_2\rangle=|\hat{A}_1=-1\rangle\otimes|\hat{B}_2=-1\rangle,\\
|\Phi_3\rangle=|\hat{A}_2=-1 \rangle\otimes|\hat{B}_1=-1\rangle,\\
|\Phi_4\rangle=|\hat{A}_2=-1\rangle\otimes|\hat{B}_2=-1\rangle.
\end{array}
\right \}
\end{equation}

\subsection{A 3-d subspace $S$}
The states $|\Phi_1\rangle$, $|\Phi_2\rangle$ and $|\Phi_3\rangle$ are linearly independent and span a 3-d subspace $S$, of ${C\!\!\!\!I}^{\ 2} \otimes {C\!\!\!\!I}^{\ 2}$.

\textbf{Proof:}
Consider a linear combination of $|\Phi_1\rangle$, $|\Phi_2\rangle$ and $|\Phi_3\rangle$ equated to zero:
\begin{eqnarray}
&&a|\Phi_1\rangle+b|\Phi_2\rangle+c|\Phi_3\rangle=0,\nonumber\\
\Rightarrow
&&a|\hat{A}_1=+1\rangle\otimes|\hat{B}_1=+1\rangle+b|\hat{A}_1=-1\rangle\otimes|\hat{B}_2=-1\rangle\nonumber\\
&&+c|\hat{A}_2=-1\rangle\otimes|\hat{B}_1=-1\rangle=0.
\end{eqnarray}
\begin{eqnarray*}
\rm{i.e.}&&aa_{11}b_{11}|\hat{A}_2=+1\rangle\otimes|\hat{B}_2=+1\rangle+(aa_{11}b_{12}+ba_{21})|\hat{A}_2=+1\rangle\otimes|\hat{B}_2=-1\rangle\\&&+(aa_{12}b_{11}+cb_{21})|\hat{A}_2=-1\rangle\otimes|\hat{B}_2=+1\rangle\\&&+(aa_{12}b_{12}+ba_{22}+cb_{22})|\hat{A}_2=-1\rangle\otimes|\hat{B}_2=-1\rangle=0.
\end{eqnarray*}

Since the states in the above equation are mutually orthogonal and all $a_{ij}$'s and $b_{ij}$'s ($i,j \ \in \{1,\ 2\}$) are strictly non-zero, there exists only a trivial solution, i.e., $a=b=c=0$. Hence, the three states span a 3-d subspace $S$, of $C\!\!\!\!I^{\ 2} \otimes C\!\!\!\!I^{\ 2}$. \textbf{QED}

It is also useful to note here that the states $|\Phi_1\rangle$, $|\Phi_2\rangle$, $|\Phi_3\rangle$ and $|\Phi_4\rangle$ can in fact be shown to be linearly independent.

\subsection{The Hardy subspace}
A state $|\psi\rangle$ that responds to the Hardy's nonlocality test, in order to satisfy the correlation conditions in Eq.(\ref{prob1by2}), needs to be present in the subspace of $C\!\!\!\!I^{\ 2} \otimes C\!\!\!\!I^{\ 2}$, which is orthogonal to $S$, call it the Hardy subspace ($S^{\independent}$). Since $S$ is a 3-d subspace, the Hardy subspace spans a single dimension. Thus there is a unique state $|\psi\rangle$ in the Hardy subspace corresponding to this system and we call it the Hardy state.

\subsection{Gram-Schmidt Orthonormalization}
Using Gram-Schmidt orthonormalization procedure, one can find an orthonormal basis $\{|\Phi'_1\rangle,\ |\Phi'_2\rangle,\ |\Phi'_3\rangle\}$ for $S$ as follows:
\begin{equation*}
\label{gs1by2}
\left.
\begin{array}{lcl}
|\Phi'_1\rangle=|\Phi_1\rangle,\\
|\Phi'_2\rangle=|\Phi_2\rangle,\\
|\Phi'_3\rangle=\frac{|\Phi_3\rangle-\langle\Phi'_2|\Phi_3\rangle|\Phi'_2\rangle}{\sqrt{1-(|\langle\Phi'_2|\Phi_3\rangle|^2)}}
\end{array}
\right\}
\end{equation*}

The Hardy state $|\psi\rangle$, will then be given by:
\begin{eqnarray}
&&|\psi\rangle=\frac{|\Phi_4\rangle-\langle\Phi'_1|\Phi_4\rangle|\Phi'_1\rangle-\langle\Phi'_2|\Phi_4\rangle|\Phi'_2\rangle-\langle\Phi'_3|\Phi_4\rangle|\Phi'_3\rangle}{\sqrt{1-(|\langle\Phi'_1|\Phi_4\rangle|^2+|\langle\Phi'_2|\Phi_4\rangle|^2+|\langle\Phi'_3|\Phi_4\rangle|^2)}}.
\label{psimax1by2}
\end{eqnarray}

The amount of nonlocality $q$, exhibited by this state, can be found to be:
\begin{eqnarray}
q&=&|\langle\psi|\Phi_4\rangle|^2,\nonumber\\
&=&1- \displaystyle\sum_{i=1}^3|w_{i}|^2,
\label{q1by2''}
\end{eqnarray}
where $w_i=\langle \Phi'_i|\Phi_4\rangle$ for $i \in \{1,\ 2,\ 3\}$.
Now, one can find out the values of $|w_{i}|^2$ in terms of the coefficients $a_{jk}$'s and $b_{jk}$'s that appeared in Eq.(\ref{eigenvec1by2}). $q$ then takes the form:
\begin{eqnarray}
q&=&(1-|a_{22}|^2)-\frac{|b_{22}|^2(1-|a_{22}|^2)^2}{(1-|b_{22}|^2|a_{22}|^2)}-|a_{12}|^2|b_{12}|^2.
\label{q1by2'}
\end{eqnarray}

Upon substitution of the values of these coefficients in terms of $\theta_1$, $\theta_2$, $\phi_1$ and $\phi_2$, $q$ takes the form:
\begin{eqnarray}
q&=&-\frac{\sin^2\theta_1\sin^2\theta_2}{4(-3+\cos\theta_1+\cos\theta_2+\cos\theta_1\cos\theta_2)}.
\label{q1by2}
\end{eqnarray}

\subsection{Maximizing $q$ through observables}
The amount of nonlocality $q$ exhibited by the Hardy state $|\psi\rangle$, can now be maximized by a suitable choice of the observables $\hat{A_1},\ \hat{A_2},\ \hat{B_1}$ and $\hat{B_2}$, i.e., by a suitable choice of $\theta_1$, $\theta_2 \in (0,\ \pi)$ (with arbitrarily chosen $\phi_1$, $\phi_2\ \in [0,\ 2\pi)$). The two equations
\begin{equation*}
\left.
\begin{array}{lcl}
\frac{\partial q}{\partial \theta_1}\equiv-\frac{(3-12\cos\theta_1+\cos2\theta_1+8\cos^4\frac{\theta_1}{2}\cos\theta_2)\sin\theta_1\sin^2\theta_2}{8(-3+\cos\theta_1+2\cos^2\frac{\theta_1}{2}\cos\theta_2)^2}=0,\\
\frac{\partial q}{\partial \theta_2}\equiv-\frac{(3-12\cos\theta_2+\cos2\theta_2+8\cos^4\frac{\theta_2}{2}\cos\theta_1)\sin\theta_2\sin^2\theta_1}{8(-3+\cos\theta_2+2\cos^2\frac{\theta_2}{2}\cos\theta_1)^2}=0,
\end{array}
\right\}
\end{equation*}
give rise to the following solution: $(\cos\theta_1,\ \cos\theta_2) = (-2+\sqrt{5},\ -2+\sqrt{5})$.

Thus, we see that the optimal value of the symmetric function $q$ in Eq.(\ref{q1by2}) occurs on the plane $\theta_1=\theta_2$. Now taking $\theta_1=\theta_2=\theta$(say) in Eq.(\ref{q1by2}) and maximizing $q$ over $\theta$, one can see that at that value of $\theta \ \in (0,\pi)$ for which $\cos\theta=-2+\sqrt{5}$, $q$ attains its maximum value, which is equal to $\frac{-11+5\sqrt{5}}{2}\approx 0.0901699$. The corresponding value of $\theta$ is $76.35$ degrees.

\subsection{Hardy's nonlocality for a given state}
\label{invariants}

Given {\it any} non-maximally entangled two-qubit pure state $|\psi'\rangle$, one can always find out the largest eigenvalue $\lambda_{\psi'} \in (\frac{1}{2},\ 1)$ of the single qubit reduced density matrix $\rho^{\psi'}_{2}=\rm{Tr_{2}}(|\psi'\rangle\langle\psi'|)$. Note that $\lambda_{\psi'}$ is the only SU($2$)$\otimes$SU($2$)-invariant of the two-qubit pure state $|\psi'\rangle$. The determinant of $\rho^{\psi'}_{2}$, $I_{\psi'}=\lambda_{\psi'} (1-\lambda_{\psi'})\in(0,\ \frac{1}{4})$, a function of $\lambda_{\psi'}$, can also be taken as the invariant quantity instead of $\lambda_{\psi'}$.

Now, consider the state $|\psi\rangle$ in Eq.(\ref{psimax1by2}). It takes the following form in the standard basis (i.e. the basis of the observables $\hat{A}_2 \equiv \hat{S_z}$ and $\hat{B}_2 \equiv \hat{S_z}$):
\begin{eqnarray}
&|\psi\rangle&=\frac{a_{11} a_{12}^*b_{11}b_{12}^*}{|a_{11}a_{12}b_{11}b_{12}|\sqrt{1-|a_{11}b_{11}|^2}}\nonumber\\
&&\times[\sqrt{1-|a_{11}b_{11}|^2-|a_{11}a_{12}b_{11}|^2}|+1\rangle\otimes\frac{(1-|a_{11}b_{11}|^2)|+1\rangle-|a_{11}|^2b_{11}^*b_{12}|-1\rangle}{\sqrt{1-|a_{11}b_{11}|^2-|a_{11}b_{11}b_{12}|^2}}\nonumber\\
&&-a_{11}^*a_{12}b_{11}^*|-1\rangle\otimes(b_{11}|+1\rangle+b_{12}|-1\rangle)].
\label{psi1std}
\end{eqnarray}

The single qubit reduced density matrix $\rho^{\psi}_{2}$, is then given by:
\begin{eqnarray}
\rho^{\psi}_{2}&=&\frac{1}{1-|a_{11}b_{11}|^2} \left( \begin{array}{cc}
(1-|a_{11}b_{11}|^2-|a_{11}b_{11}b_{12}|^2) & -b_{11}b_{12}^*|a_{11}b_{12}|^2 \\
-b_{11}^*b_{12}|a_{11}b_{12}|^2 & |a_{11}b_{11}b_{12}|^2  \end{array} \right),
\label{rhopsi}
\end{eqnarray}
whose characteristic equation can be found out to be:
\begin{eqnarray}
\lambda^2-\lambda+\frac{|a_{11}a_{12}b_{11}b_{12}|^2}{(1-|a_{11}b_{11}|^2)^2}&=&0.
\label{chareq}
\end{eqnarray}

Hence, 
\begin{eqnarray}
I_{\psi}&=&\frac{|a_{11}a_{12}b_{11}b_{12}|^2}{(1-|a_{11}b_{11}|^2)^2}\nonumber\\
&=&\frac{\sin^2\theta_1\sin^2\theta_2}{(3-\cos\theta_1-\cos\theta_2-\cos\theta_1\cos\theta_2)^2} \ \in(0,\frac{1}{4}).
\label{invhalf}
\end{eqnarray}

By varying the values of the four parameters $\theta_1$, $\theta_2$, $\phi_1$, $\phi_2$ over all possible allowed values, one can check that $I_{\psi}$ will scan the entire interval $(0,\frac{1}{4})$. Hence, there will exist at least one set of values ($\theta_1,\ \theta_2,\ \phi_1,\ \phi_2$), for which $I_{\psi}$=$I_{\psi'}$. This will immediately imply that $|\psi'\rangle$ is locally unitarily connected to $|\psi\rangle$ for the above-mentioned choice of ($\theta_1,\ \theta_2,\ \phi_1,\ \phi_2$). So, $|\psi'\rangle$ will also satisfy the Hardy's nonlocality conditions. That any given non-maximally entangled two qubit pure state satisfies the Hardy's nonlocality conditions, was shown earlier by Goldstein \cite{gold}.

\section{A system of two spin-$1$ particles}
\label{spin1}
The spin-$1$ observables $\hat{A}_1$ and $\hat{B}_1$ may be written in their most general form, in the eigen-bases of $\hat{A}_2 \equiv \hat{S_z}$ and $\hat{B}_2 \equiv \hat{S_z}$, as:
\[\hat{A}_1=\left( \begin{array}{ccc}
\cos\theta_1 &\frac{e^{-i\phi_1}\sin\theta_1}{\sqrt2}  & 0 \\
\frac{e^{i\phi_1}\sin\theta_1}{\sqrt2} & 0 & \frac{e^{-i\phi_1}\sin\theta_1}{\sqrt2} \\
0 & \frac{e^{i\phi_1}\sin\theta_1}{\sqrt2} & -\cos\theta_1 \end{array} \right) \] \
and a similar expression for $\hat{B}_1$ with $(\theta_1,\ \phi_1)$ replaced by $(\theta_2,\ \phi_2)$.

Their eigenvectors are then given by:
\begin{equation}
\label{eigenvec1}
\left.
\begin{array}{lcl}
|\hat{A}_1=+1\rangle=a_{11}|+1\rangle +a_{12}|0\rangle +a_{13}|-1\rangle,\\
|\hat{A}_1=0\rangle=a_{21}|+1\rangle +a_{22}|0\rangle +a_{23}|-1\rangle,\\
|\hat{A}_1=-1\rangle=a_{31}|+1\rangle +a_{32}|0\rangle +a_{33}|-1\rangle,\\
|\hat{B}_1=+1\rangle=b_{11}|+1\rangle +b_{12}|0\rangle +b_{13}|-1\rangle,\\
|\hat{B}_1=0\rangle=b_{21}|+1\rangle +b_{22}|0\rangle +b_{23}|-1\rangle,\\
|\hat{B}_1=-1\rangle=b_{31}|+1\rangle +b_{32}|0\rangle +b_{33}|-1\rangle,\\
\end{array}
\right \}
\end{equation}
where
\[\left( \begin{array}{ccc}
a_{11} & a_{12} & a_{13} \\
a_{21} & a_{22} & a_{23} \\
a_{31} & a_{32} & a_{33}  \end{array} \right) =\left( \begin{array}{ccc}
\frac{e^{-2 i \phi_1} (1+\cos\theta_1)}{2} & \frac{e^{-i \phi_1} \sin\theta_1}{\sqrt{2}} & \frac{(1-\cos\theta_1)}{2} \\
-\frac{e^{-2 i \phi_1} \sin\theta_1}{\sqrt{2}} & e^{-i \phi_1} \cos\theta_1 & \frac{\sin\theta_1}{ \sqrt{2}} \\
\frac{e^{-2 i \phi_1} (1-\cos\theta_1)}{2} & -\frac{e^{-i \phi_1} \sin\theta_1}{\sqrt{2}} & \frac{1+\cos\theta_1}{2}  \end{array} \right)\]
and similar expressions for $b_{ij}$'s with $(\theta_1,\ \phi_1)$ replaced by $(\theta_2,\ \phi_2)$. In this system, the non-commutativity of observables at each observer gives rise to the following distinct possibilities with respect to the relationship between their eigenvectors. The eigenvectors of the two observables (say $\hat{X}_1$ and $\hat{X}_2$) at an observer could be such that:

\begin{enumerate}
\item $0<|\langle\hat{X_1}=i|\hat{X_2}=j\rangle|<1$, for all $i,j \ \in \{+1,\ 0,\ -1\}$ (or)
\item $|\langle\hat{X_1}=i|\hat{X_2}=j\rangle|=1$ for a single pair ($i,j$), where $i,j\ \in \{+1,\ 0,\ -1\}$ (or)
\item $|\langle\hat{X_1}=i|\hat{X_2}=j\rangle|=0$ for a single pair ($i,j$), where $i,j\ \in \{+1,\ 0,\ -1\}$.
\end{enumerate}

However, considering the fact that the observables are spin-$1$ observables, the $2^{\rm{nd}}$ case turns out to be infeasible, while the $3^{\rm{rd}}$ case occurs only at $\theta_1=\frac{\pi}{2}$ (and $\theta_2=\frac{\pi}{2}$), making $a_{22}$ (and $b_{22}$) zero. Without any loss of generality, we assume the $1^{\rm{st}}$ case to hold good in the following analysis (the case $\theta_1=\theta_2=\frac{\pi}{2}$ will be automatically included in this analysis).

\subsection{The Hardy's nonlocality test}

The following conditions constitute the Hardy's nonlocality test for the given system, in it's minimal form:
\begin{equation}
\label{prob1}
\left.
\begin{array}{lcl}
\rm prob(\hat{A}_1=+1,\ \hat{B_1}=+1)=0, \\
\rm prob(\hat{A}_1=0,\ \hat{B_2}=-1)=0, \\
\rm prob(\hat{A}_1=-1,\ \hat{B_2}=-1)=0, \\
\rm prob(\hat{A}_2=-1,\ \hat{B_1}=0)=0, \\
\rm prob(\hat{A}_2=-1,\ \hat{B_1}=-1)=0, \\
\rm prob(\hat{A}_2=-1,\ \hat{B_2}=-1)=q \ > 0.
\end{array}
\right \}
\end{equation}

We associate a state with every condition in the test, say:
\begin{equation}
\label{ket1}
\left.
\begin{array}{lcl}
|\Phi_1\rangle=|\hat{A}_1=+1\rangle\otimes|\hat{B}_1=+1\rangle,\\
|\Phi_2\rangle=|\hat{A}_1=0 \rangle\otimes|\hat{B}_2=-1\rangle,\\
|\Phi_3\rangle=|\hat{A}_1=-1\rangle\otimes|\hat{B}_2=-1\rangle,\\
|\Phi_4\rangle=|\hat{A}_2=-1\rangle\otimes|\hat{B}_1=0\rangle,\\
|\Phi_5\rangle=|\hat{A}_2=-1\rangle\otimes|\hat{B}_1=-1\rangle,\\
|\Phi_6\rangle=|\hat{A}_2=-1\rangle\otimes|\hat{B}_2=-1\rangle.
\end{array}
\right \}
\end{equation}

\subsection{A 5-d subspace $S$}
\label{summa1}
The states $|\Phi_1\rangle$, $|\Phi_2\rangle$, $|\Phi_3\rangle$, $|\Phi_4\rangle$ and $|\Phi_5\rangle$ are linearly independent and span a 5-d subspace $S$, of $C\!\!\!\!I^{\ 3} \otimes C\!\!\!\!I^{\ 3}$. The proof for this, can be found in Appendix A. It is also useful to note here that the states $|\Phi_1\rangle$, $|\Phi_2\rangle$, $|\Phi_3\rangle$, $|\Phi_4\rangle$, $|\Phi_5\rangle$ and $|\Phi_6\rangle$ can in fact be shown to be linearly independent.

\subsection{A 4-d subspace $S^{\independent}$}

Consider the subspace of $C\!\!\!\!I^{\ 3} \otimes C\!\!\!\!I^{\ 3}$, which is orthogonal to $S$. It's a 4-d subspace, call it $S^{\independent}$. We also call it the Hardy subspace. We now look for all the product states in $S^{\independent}$. Let us choose $|\eta\rangle$ to be orthogonal to $|\hat{B_1}=+1\rangle$ and $|\hat{B_2}=-1\rangle$ (Note that by our assumption, $0<|\langle\hat{B_1}=+1|\hat{B_2}=-1\rangle|<1$.), then  $|\phi\rangle\otimes|\eta\rangle$ is orthogonal to $|\Phi_1\rangle$, $|\Phi_2\rangle$ and $|\Phi_3\rangle$, irrespective of $|\phi\rangle$. Now, by taking $|\phi\rangle$=$\lambda|\hat{A}_2=+1\rangle+\mu|\hat{A}_2=0\rangle$, it is ensured that $|\phi\rangle\otimes|\eta\rangle$ is orthogonal to $|\Phi_4\rangle$ and $|\Phi_5\rangle$. Similarly, taking $|\phi\rangle$ to be orthogonal to $|\hat{A}_2=-1\rangle$ and $|\hat{A}_1=+1\rangle$, $|\eta\rangle$ takes the form: $|\eta\rangle$=$\lambda'|\hat{B_2}=+1\rangle+\mu'|\hat{B_2}=0\rangle$. Hence, one can write four product states that are orthogonal to the subspace $S$, which are as follows:
\begin{equation}
\label{pdt}
\left.
\begin{array}{lcl}
|\Phi_7\rangle=|\hat{A}_2=+1\rangle\otimes|\eta\rangle,\\
|\Phi_8\rangle=|\hat{A}_2=0\rangle\otimes|\eta\rangle,\\
|\Phi_9\rangle=|\phi\rangle\otimes|\hat{B_2}=+1\rangle,\\
|\Phi_{10}\rangle=|\phi\rangle\otimes|\hat{B_2}=0\rangle,
\end{array}
\right \}
\end{equation}
where $|\eta\rangle$ is orthogonal to both $|\hat{B}_1=+1\rangle$ and $|\hat{B}_2=-1\rangle$ and $|\phi\rangle$ is orthogonal to $|\hat{A}_1=+1\rangle$ and $|\hat{A}_2=-1\rangle$.

In fact, the states $|\Phi_7\rangle$, $|\Phi_8\rangle$, $|\Phi_9\rangle$ and $|\Phi_{10}\rangle$ are the only four product states in the $C\!\!\!\!I^{\ 3} \otimes C\!\!\!\!I^{\ 3}$ space that are orthogonal to the subspace $S$.

\textbf{Proof:} Let $|\chi\rangle\otimes|\zeta\rangle$ be any product state of the system other than the states mentioned in Eq.(\ref{pdt}), which is orthogonal to S, i.e., orthogonal to the states $|\Phi_1\rangle, |\Phi_2\rangle, |\Phi_3\rangle, |\Phi_4\rangle$ and $|\Phi_5\rangle$ of Eq.(\ref{ket1}). Then, it has to satisfy the following conditions:
\begin{eqnarray*}
&&(\rm{either}\ \langle\chi|\hat{A}_1=+1\rangle=0\ \rm{or}\ \langle\zeta|\hat{B}_1=+1\rangle=0 \ \rm{or \ both}), \\
&&(\rm{either}\ \langle\chi|\hat{A}_1=0\rangle=0\ \rm{or}\ \langle\zeta|\hat{B}_2=-1\rangle=0 \ \rm{or \ both}),\\
&&(\rm{either}\ \langle\chi|\hat{A}_1=-1\rangle=0\ \rm{or}\ \langle\zeta|\hat{B}_2=-1\rangle=0 \ \rm{or \ both}),\\
&&(\rm{either}\ \langle\chi|\hat{A}_2=-1\rangle=0\ \rm{or}\ \langle\zeta|\hat{B}_1=0\rangle=0 \ \rm{or \ both}),\\
&&(\rm{either}\ \langle\chi|\hat{A}_2=-1\rangle=0\ \rm{or}\ \langle\zeta|\hat{B}_1=-1\rangle=0 \ \rm{or \ both}).\\
\end{eqnarray*}
There will be at least $2^5$ conditions--not all of them are feasible (for example, $\langle\chi|\hat{A}_1=+1\rangle=\langle\chi|\hat{A}_1=0\rangle=\langle\chi|\hat{A}_1=-1\rangle=\langle\zeta|\hat{B}_1=0\rangle=\langle\zeta|\hat{B}_1=-1\rangle=0$ is not possible). We now enumerate all the apparently feasible conditions. For example, $\langle\chi|\hat{A}_1=+1\rangle=\langle\chi|\hat{A}_1=0\rangle=0$. This will give us $|\chi\rangle=|\hat{A}_1=-1\rangle$, hence $\langle\zeta|\hat{B}_2=-1\rangle=0$. However, $\langle\chi|\hat{A}_2=-1\rangle=\langle\hat{A}_1=-1|\hat{A}_2=-1\rangle\not=0$, since $0<|\langle\hat{A_1}=-1|\hat{A_2}=-1\rangle|<1$, which immediately implies that $\langle\zeta|\hat{B}_1=0\rangle=\langle\zeta|\hat{B}_1=-1\rangle=0$. So, we must have $|\zeta\rangle=|\hat{B}_1=+1\rangle$. However, $\langle\zeta|\hat{B}_2=-1\rangle=\langle\hat{B}_1=+1|\hat{B}_2=-1\rangle\not=0$, since $0<|\langle\hat{B_1}=+1|\hat{B_2}=-1\rangle|<1$ (according to the initial assumption). Hence, we have reached at a contradiction. Similar types of contradiction occur in all such apparently feasible choices other than the states $|\Phi_7\rangle, |\Phi_8\rangle, |\Phi_9\rangle, |\Phi_{10}\rangle$. \textbf{QED}

\subsection{Forms of $|\eta\rangle$ and $|\phi\rangle$}
The forms of the states $|\eta\rangle$ and $|\phi\rangle$ can be obtained by using the orthogonality conditions imposed on them. Since $|\eta\rangle$ is orthogonal to $|\hat{B_1}=+1\rangle$ and $|\hat{B_2}=-1\rangle$, it may be written as:
\begin{eqnarray}
&&|\eta\rangle=\alpha|\hat{B_2}=+1\rangle+\beta|\hat{B_2}=0\rangle,\nonumber\\
\rm{with}\ &&\langle\hat{B_1}=+1|\eta\rangle=0.
\label{inn1}
\end{eqnarray}

Substituting $|\hat{B_1}=+1\rangle=b_{11}|\hat{B_2}=+1\rangle+b_{12}|\hat{B_2}=0\rangle+b_{13}|\hat{B_2}=-1\rangle$ in Eq.(\ref{inn1}), while $|b_{11}|^2+|b_{12}|^2+|b_{13}|^2=1$ and $0<|\langle\hat{B_1}=+1|\hat{B_2}=j\rangle|<1$, for all $j \ \in \{+1,\ 0,\ -1\}$, we get:
\begin{eqnarray*}
b_{11}^*\alpha+b_{12}^*\beta&=&0, \nonumber\\
\Rightarrow\ \beta&=&-\frac{b_{11}^*}{b_{12}^*}\alpha \ \rm{(as\ b_{12}\not=0)}.
\end{eqnarray*}

Hence, $|\eta\rangle$ can be written as:
\begin{eqnarray*}
|\eta\rangle&=&\alpha|\hat{B_2}=+1\rangle-\frac{b_{11}^*}{b_{12}^*}\alpha|\hat{B_2}=0\rangle.
\end{eqnarray*}

Taking $\alpha$=$b_{12}^*$ and normalizing the state, we get:
\begin{eqnarray}
|\eta\rangle&=&\frac{b_{12}^*|\hat{B_2}=+1\rangle-b_{11}^*|\hat{B_2}=0\rangle}{\sqrt{|b_{11}|^2+|b_{12}|^2}}.
\end{eqnarray}

On similar lines, $|\phi\rangle$ can be found to be:
\begin{eqnarray}
|\phi\rangle&=&\frac{a_{12}^*|\hat{A}_2=+1\rangle-a_{11}^*|\hat{A}_2=0\rangle}{\sqrt{|a_{11}|^2+|a_{12}|^2}}.
\end{eqnarray}

\subsection{A 3-d subspace $S'$}
\label{sprime1}
 The states $|\Phi_7\rangle,|\Phi_8\rangle,|\Phi_9\rangle$ and $|\Phi_{10}\rangle$ span a 3-d subspace $S'$, of $S^{\independent}$. The proof for this, can be found in Appendix B. Let us choose the set of states $\{|\Phi_i\rangle,\ i:7 \rm{\ to\ }9 \}$ to span $S'$. Moreover, one must also note that the states $|\Phi_7\rangle, |\Phi_8\rangle, |\Phi_9\rangle$ and $|\Phi_{10}\rangle$ are orthogonal not only to the states $|\Phi_1\rangle, |\Phi_2\rangle, |\Phi_3\rangle, |\Phi_4\rangle$ and $|\Phi_{5}\rangle$, but also to the state $|\Phi_6\rangle$.

\subsection{The Hardy subspace}
\label{mostgen1}
The most general state $|\Psi\rangle$ satisfying the conditions in Eq.(\ref{prob1}), has to be of the form $|\Psi\rangle=v_{0}|\psi\rangle+\displaystyle\sum_{i=7}^9v_{i}|\Phi_{i}\rangle$, with $\displaystyle\sum_{i,\ j=7}^9v_{i}v_{j}^*\langle\Phi_{j}|\Phi_{i}\rangle+|v_0|^2=1,\ v_0\not=0,$ where $|\psi\rangle$ spans the one-dimensional subspace $(S\oplus S')^{\independent}$ of the Hardy subspace $S^{\independent}$. Using Gram-Schmidt orthonormalization method, one can get the unique form of $|\psi\rangle$ as:
\begin{eqnarray}
&&|\psi\rangle=\frac{|\Phi_6\rangle-\langle\Phi'_1|\Phi_6\rangle|\Phi'_1\rangle-\langle\Phi'_2|\Phi_6\rangle|\Phi'_2\rangle-\langle\Phi'_3|\Phi_6\rangle|\Phi'_3\rangle-\langle\Phi'_4|\Phi_6\rangle|\Phi'_4\rangle-\langle\Phi'_5|\Phi_6\rangle|\Phi'_5\rangle}{\sqrt{1-(|\langle\Phi'_1|\Phi_6\rangle|^2+|\langle\Phi'_2|\Phi_6\rangle|^2+|\langle\Phi'_3|\Phi_6\rangle|^2+|\langle\Phi'_4|\Phi_6\rangle|^2+|\langle\Phi'_5|\Phi_6\rangle|^2)}},\nonumber\\
\label{psimax1}
\end{eqnarray}
with $|\Phi'_1\rangle$, $|\Phi'_2\rangle$, $|\Phi'_3\rangle$, $|\Phi'_4\rangle$ and $|\Phi'_5\rangle$ given by:
\begin{equation*}
\label{gs1}
\left.
\begin{array}{lcl}
|\Phi'_1\rangle=|\Phi_1\rangle,\\
|\Phi'_2\rangle=|\Phi_2\rangle,\\
|\Phi'_3\rangle=|\Phi_3\rangle,\\
|\Phi'_4\rangle=\frac{|\Phi_4\rangle-\langle\Phi_2|\Phi_4\rangle|\Phi_2\rangle-\langle\Phi_3|\Phi_4\rangle|\Phi_3\rangle}{\sqrt{1-(|\langle\Phi_2|\Phi_4\rangle|^2+|\langle\Phi_3|\Phi_4\rangle|^2)}},\\
|\Phi'_5\rangle=\frac{|\Phi''_5\rangle-\langle\Phi'_4|\Phi''_5\rangle|\Phi'_4\rangle}{\sqrt{1-|\langle\Phi'_4|\Phi''_5\rangle|^2}},
\end{array}
\right\}
\end{equation*}
\begin{eqnarray*}
\rm{where\ }&&|\Phi''_5\rangle=\frac{|\Phi_5\rangle-\langle\Phi_2|\Phi_5\rangle|\Phi_2\rangle-\langle\Phi_3|\Phi_5\rangle|\Phi_3\rangle}{\sqrt{1-(|\langle\Phi_2|\Phi_5\rangle|^2+|\langle\Phi_3|\Phi_5\rangle|^2)}}.
\label{gs2}
\end{eqnarray*}

\subsection{Functional form of the amount of nonlocality $q$}
\label{func1}
The amount of nonlocality $q$ for the Hardy state $|\Psi\rangle=|\psi\rangle$, can be seen to be the maximum amongst all possible Hardy states, for a given choice of observables; and is given by:
\begin{eqnarray}
q&=&|\langle\psi|\Phi_6\rangle|^2\nonumber\\
&=&1-\displaystyle\sum_{i=1}^5|w_{i}|^2,
\label{gen7}
\end{eqnarray}
where $w_i=\langle \Phi'_i|\Phi_6\rangle$ for $i\ \in \{1,\ 2,\ 3,\ 4,\ 5\}$. We call $|\psi\rangle$, the maximally nonlocal Hardy state of the system.

Now one can find out the values of $|w_{i}|^2$ in terms of the coefficients $a_{jk}$'s and $b_{jk}$'s that appeared in Eq.(\ref{eigenvec1}). $q$ then takes the form:
\begin{equation}
\label{q1}
\left.
\begin{array}{lcl}
q=(1-(|a_{23}|^2+|a_{33}|^2))-\frac{|b_{23}|^2(1-(|a_{23}|^2+|a_{33}|^2))^2}{(1-|b_{23}|^2(|a_{23}|^2+|a_{33}|^2))}\\
-\frac{|b_{33}|^2(1-(|a_{23}|^2+|a_{33}|^2))^2}{(1-(|b_{23}|^2+|b_{33}|^2)(|a_{23}|^2+|a_{33}|^2))(1-|b_{23}|^2(|a_{23}|^2+|a_{33}|^2))}-|a_{13}b_{13}|^2.
\end{array}
\right\}
\end{equation}

Upon substitution of the coefficients in terms of $\theta_1$, $\theta_2$, $\phi_1$ and $\phi_2$, $q$ takes the form:
\begin{eqnarray}
q&=& \frac{16\cos^2\frac{\theta_1}{2}\cos^2\frac{\theta_2}{2}(-3+\cos\theta_1)(-3+\cos\theta_2)\sin^4\frac{\theta_1}{2}\sin^4\frac{\theta_2}{2}}{39-20\cos\theta_1+5\cos2\theta_1+16\cos^2\frac{\theta_1}{2}(-3+\cos\theta_1)\cos\theta_2-4\cos^2\frac{\theta_1}{2}(-3+\cos\theta_1)\cos2\theta_2}.\nonumber\\
\label{q1'}
\end{eqnarray}

\subsection{Maximizing $q$ through observables}
\label{summa3}
The amount of nonlocality $q$ exhibited by the maximally nonlocal Hardy state $|\psi\rangle$, can now be maximized by a suitable choice of observables $\hat{A_1},\ \hat{A_2},\ \hat{B_1}$ and $\hat{B_2}$, i.e., by a suitable choice of $\theta_1,\ \theta_2 \in (0,\ \pi)$. The partial derivatives of $q$ with respect to $\theta_1$ and $\theta_2$ are calculated, $\cos\theta_1$ replaced with $a$ and $\cos\theta_2$ replaced with $b$ ($a,\ b\ \in (-1,\ +1)$) and the critical points are obtained by solving the following simultaneous equations:
\begin{equation*}
\left.
\begin{array}{lcl}
\frac{\partial q}{\partial \theta_1}=0,\ \frac{\partial q}{\partial \theta_2}=0.\\
\rm{i.e.}\\
(-1+a)\sqrt{1-a^2}(1+(-2+a)a(-3+b)-3 b)(-3+b)(-1+b)^2(1+b)\\
\times\frac{(5-3b+(-2+a)a(1+b))}{8(7+3(-2+b)b+2a(-3+b)(1+b)-a^2(-3+b)(1+b))^2}=0,\\
(-3+a)(-1+a)^2(1+a)(-1+b)\sqrt{1-b^2}(1-3(-2+b)b+a(-3+b)(1+b))\\
\times\frac{(5+(-2+b)b+a(-3+b)(1+b))}{8(7+3(-2+b)b+2a(-3+b)(1+b)-a^2(-3+b)(1+b))^2}=0.
\end{array}
\right\}
\end{equation*}
The above equations give rise to the following solution: $(\cos\theta_1,\ \cos\theta_2) = (2-\sqrt{5},\ 2-\sqrt{5})$.

Thus we see that the optimal value of the symmetric function $q$ in Eq.(\ref{q1}) occurs on the plane $\theta_1=\theta_2$. Now taking $\theta_1=\theta_2=\theta$(say) in Eq.(\ref{q1}) and maximizing $q$ over $\theta$, one can see that at that value of $\theta \ \in (0,\ \pi)$ for which $\cos\theta=2-\sqrt{5}$, $q$ attains its maximum value, which is equal to $\frac{-11+5\sqrt{5}}{2}\cong 0.0901699$, and the corresponding value of $\theta$ is $103.65$ degrees.

It is to be mentioned here that in an earlier work \cite{ghosh}, Ghosh and Kar have found a value of $q$ in Hardy-type nonlocality (not in it's minimal form as given in Eq.(\ref{prob1})) argument for two spin-$1$ particles as $0.132$ (approximately). The result in the present work hints at some errors in the calculation of the maximum value of $q$ in the above-mentioned work of Ghosh and Kar, as the Hardy's nonlocality conditions used in \cite{ghosh} put further restriction on the Hardy subspace.

\subsection{Hardy's nonlocality for a given state}
Given {\it any} non-maximally entangled pure state $|\psi'\rangle$ of two spin-$1$ particles, one can find out the full set of SU(3)$\otimes$SU(3)-invariants, namely, $I_1=\lambda_1 \lambda_2 + \lambda_2 \lambda_3 + \lambda_3 \lambda_1 \ \in(0,\ \frac{1}{3})$ and $I_2=\lambda_1 \lambda_2 \lambda_3 \ \in[0,\ \frac{1}{27})$, where $\lambda_i, \ i \in \{1,\ 2,\ 3\}$ are eigenvalues of the single spin-$1$ particle reduced density matrix $\rho^{\psi'}_{2}=\rm{Tr_{2}}(|\psi'\rangle\langle\psi'|)$. For the state $|\psi\rangle$ in Eq.(\ref{psimax1}), one can find out $I_1$ and $I_2$. They will, in general be functions of $\theta_1$, $\theta_2$, $\phi_1$, $\phi_2$. However, unlike in the case of the system of two qubits, varying the values of these four parameters over all possible allowed values will not scan the entire intervals of the invariants $I_1$, $I_2$. This is attributed to the presence of the subspace $S'$ (section \ref{sprime1}) in this system which doesn't exist in the case of the two-qubit system. Nevertheless, if one considers the most general Hardy state of this system, namely, $|\Psi\rangle=v_{0}|\psi\rangle+\displaystyle\sum_{i=7}^9v_{i}|\Phi_{i}\rangle$, with $\displaystyle\sum_{i,\ j=7}^9v_{i}v_{j}^*\langle\Phi_{j}|\Phi_{i}\rangle+|v_0|^2=1,\ v_0\not=0,$ where $|\psi\rangle$ is the maximally nonlocal Hardy state (section \ref{mostgen1}), it can be shown that it's invariants scan the entire intervals $(0,\ \frac{1}{3})$ and $[0,\ \frac{1}{27})$. Hence, one can show that $|\psi'\rangle$ is locally unitarily connected to one of the Hardy states and thus satisfies the Hardy's nonlocality conditions.

\section{A spin-$\frac{3}{2}$ bipartite system}
\label{spin3by2}
The spin-$\frac{3}{2}$ observables $\hat{A}_1$ and $\hat{B}_1$ may be written in their most general form, in the eigen-bases of $\hat{A}_2 \equiv \hat{S_z}$ and $\hat{B}_2 \equiv \hat{S_z}$, as:
\[\hat{A}_1=\left( \begin{array}{cccc}
3\cos\theta_1 & \sqrt{3}\sin\theta_1 e^{-i\phi_1} & 0 & 0\\
\sqrt{3}\sin\theta_1 e^{i\phi_1} & \cos\theta_1 & 2\sin\theta_1 e^{-i\phi_1} &0 \\
0 & 2\sin\theta_1 e^{i\phi_1} & -\cos\theta_1 & \sqrt{3}\sin\theta_1 e^{-i\phi_1}\\
0 & 0 & \sqrt{3}\sin\theta_1 e^{i\phi_1} & -3\cos\theta_1 \end{array} \right)\] \
and a similar expression for $\hat{B}_1$ with $(\theta_1,\ \phi_1)$ replaced by $(\theta_2,\ \phi_2)$.

Their eigenvectors are then given by:
\begin{equation}
\label{eigenvec3by2}
\left.
\begin{array}{lcl}
|\hat{A}_1=+\frac{3}{2}\rangle=a_{11}|+\frac{3}{2}\rangle +a_{12}|+\frac{1}{2}\rangle +a_{13}|-\frac{1}{2}\rangle+a_{14}|-\frac{3}{2}\rangle,\\
|\hat{A}_1=+\frac{1}{2}\rangle=a_{21}|+\frac{3}{2}\rangle +a_{22}|+\frac{1}{2}\rangle +a_{23}|-\frac{1}{2}\rangle+a_{24}|-\frac{3}{2}\rangle,\\
|\hat{A}_1=-\frac{1}{2}\rangle=a_{31}|+\frac{3}{2}\rangle +a_{32}|+\frac{1}{2}\rangle +a_{33}|-\frac{1}{2}\rangle+a_{34}|-\frac{3}{2}\rangle,\\
|\hat{A}_1=-\frac{3}{2}\rangle=a_{41}|+\frac{3}{2}\rangle +a_{42}|+\frac{1}{2}\rangle +a_{43}|-\frac{1}{2}\rangle+a_{44}|-\frac{3}{2}\rangle,\\
|\hat{B}_1=+\frac{3}{2}\rangle=b_{11}|+\frac{3}{2}\rangle +b_{12}|+\frac{1}{2}\rangle +b_{13}|-\frac{1}{2}\rangle+b_{14}|-\frac{3}{2}\rangle,\\
|\hat{B}_1=+\frac{1}{2}\rangle=b_{21}|+\frac{3}{2}\rangle +b_{22}|+\frac{1}{2}\rangle +b_{23}|-\frac{1}{2}\rangle+b_{24}|-\frac{3}{2}\rangle,\\
|\hat{B}_1=-\frac{1}{2}\rangle=b_{31}|+\frac{3}{2}\rangle +b_{32}|+\frac{1}{2}\rangle +b_{33}|-\frac{1}{2}\rangle+b_{34}|-\frac{3}{2}\rangle,\\
|\hat{B}_1=-\frac{3}{2}\rangle=b_{41}|+\frac{3}{2}\rangle +b_{42}|+\frac{1}{2}\rangle +b_{43}|-\frac{1}{2}\rangle+b_{44}|-\frac{3}{2}\rangle,
\end{array}
\right \}
\end{equation}
where the coefficients $a_{ij}$'s and $b_{ij}$'s are given in Appendix C.

In this system, like in the case of a system of two spin-$1$ particles (section \ref{spin1}), without any loss of generality, we assume that $0<|\langle\hat{A_1}=i|\hat{A_2}=j\rangle|<1$, $0<|\langle\hat{B_1}=i|\hat{B_2}=j\rangle|<1$ for all $i,j \ \in \{+\frac{3}{2},\ +\frac{1}{2},\ -\frac{1}{2},\ -\frac{3}{2}\}$.

\subsection{The Hardy's nonlocality test}
The following conditions constitute the Hardy's nonlocality test for the given system, in it's minimal form:
\begin{equation}
\label{prob3by2}
\left.
\begin{array}{lcl}
\rm prob(\hat{A}_1=+\frac{3}{2},\ \hat{B_1}=+\frac{3}{2})=0,\\
\rm prob(\hat{A}_1=+\frac{1}{2},\ \hat{B_2}=-\frac{3}{2})=0,\\
\rm prob(\hat{A}_1=-\frac{1}{2},\ \hat{B_2}=-\frac{3}{2})=0,\\
\rm prob(\hat{A}_1=-\frac{3}{2},\ \hat{B_2}=-\frac{3}{2})=0,\\
\rm prob(\hat{A}_2=-\frac{3}{2},\ \hat{B_1}=+\frac{1}{2})=0,\\
\rm prob(\hat{A}_2=-\frac{3}{2},\ \hat{B_1}=-\frac{1}{2})=0,\\
\rm prob(\hat{A}_2=-\frac{3}{2},\ \hat{B_1}=-\frac{3}{2})=0,\\
\rm prob(\hat{A}_2=-\frac{3}{2},\ \hat{B_2}=-\frac{3}{2})=q \ > 0.
\end{array}
\right \}
\end{equation}

We associate a state with every condition in the test, say:
\begin{equation}
\label{ket3by2}
\left.
\begin{array}{lcl}
|\Phi_1\rangle=|\hat{A}_1=+\frac{3}{2}\rangle\otimes|\hat{B}_1=+\frac{3}{2}\rangle,\\
|\Phi_2\rangle=|\hat{A}_1=+\frac{1}{2} \rangle\otimes|\hat{B}_2=-\frac{3}{2}\rangle,\\
|\Phi_3\rangle=|\hat{A}_1=-\frac{1}{2}\rangle\otimes|\hat{B}_2=-\frac{3}{2}\rangle,\\
|\Phi_4\rangle=|\hat{A}_1=-\frac{3}{2}\rangle\otimes|\hat{B}_2=-\frac{3}{2}\rangle,\\
|\Phi_5\rangle=|\hat{A}_2=-\frac{3}{2}\rangle\otimes|\hat{B}_1=+\frac{1}{2}\rangle,\\
|\Phi_6\rangle=|\hat{A}_2=-\frac{3}{2}\rangle\otimes|\hat{B}_1=-\frac{1}{2}\rangle,\\
|\Phi_7\rangle=|\hat{A}_2=-\frac{3}{2} \rangle\otimes|\hat{B}_1=-\frac{3}{2}\rangle,\\
|\Phi_8\rangle=|\hat{A}_2=-\frac{3}{2}\rangle\otimes|\hat{B}_2=-\frac{3}{2}\rangle.
\end{array}
\right \}
\end{equation}

\subsection{A 7-d subspace $S$}

As in subsection $\ref{summa1}$, it can be shown that the states $|\Phi_1\rangle$, $|\Phi_2\rangle$, $|\Phi_3\rangle$, $|\Phi_4\rangle$, $|\Phi_5\rangle$, $|\Phi_6\rangle$ and $|\Phi_7\rangle$ are linearly independent and span a 7-d subspace $S$, of $C\!\!\!\!I^{\ 4} \otimes C\!\!\!\!I^{\ 4}$. It is also useful to note here that the states $|\Phi_1\rangle$, $|\Phi_2\rangle$, $|\Phi_3\rangle$, $|\Phi_4\rangle$, $|\Phi_5\rangle$, $|\Phi_6\rangle$, $|\Phi_7\rangle$ and $|\Phi_8\rangle$ can in fact be shown to be linearly independent.

\subsection{A 9-d subspace $S^{\independent}$}

Consider the subspace of $C\!\!\!\!I^{\ 4} \otimes C\!\!\!\!I^{\ 4}$, which is orthogonal to $S$. It's a 9-d subspace, call it $S^{\independent}$. We also call it the Hardy subspace. We now look for all product states in $S^{\independent}$. Let us choose two orthogonal states $|\chi_1\rangle$ and $|\chi_2\rangle$, both being orthogonal to $|\hat{A_1}=+\frac{3}{2}\rangle$ and $|\hat{A_2}=-\frac{3}{2}\rangle$. They must satisfy the following conditions:
\begin{eqnarray*}
&&|\chi_i\rangle=a_{i}|\hat{A}_2=+\frac{3}{2}\rangle+b_{i}|\hat{A}_2=+\frac{1}{2}\rangle+c_{i}|\hat{A}_2=-\frac{1}{2}\rangle,\\
&&\ \rm{at\ least\ two\ of}\ a_{i},\ b_{i},\ c_{i}\not=0\ \rm{for\ a\ given\ i},\  i\ \in \{1,\ 2\},\\
\rm{and} \ && a_{1}^*a_{2}+b_{1}^*b_{2}+c_{1}^*c_{2}=0.
\end{eqnarray*}

We now consider the following product states:
\begin{equation}
\label{prod3by21}
\left.
\begin{array}{lcl}
|\Phi_{9}\rangle=|\chi_1\rangle\otimes|\hat{B}_2=+\frac{3}{2}\rangle,\\
|\Phi_{10}\rangle=|\chi_2\rangle\otimes|\hat{B}_2=+\frac{3}{2}\rangle,\\
|\Phi_{11}\rangle=|\chi_1\rangle\otimes|\hat{B}_2=+\frac{1}{2}\rangle,\\
|\Phi_{12}\rangle=|\chi_2\rangle\otimes|\hat{B}_2=+\frac{1}{2}\rangle,\\
|\Phi_{13}\rangle=|\chi_1\rangle\otimes|\hat{B}_2=-\frac{1}{2}\rangle,\\
|\Phi_{14}\rangle=|\chi_2\rangle\otimes|\hat{B}_2=-\frac{1}{2}\rangle.
\end{array}
\right \}
\end{equation}

Similarly, let us choose two orthogonal states $|\eta_1\rangle$ and $|\eta_2\rangle$, both being orthogonal to $|\hat{B_1}=+\frac{3}{2}\rangle$ and $|\hat{B_2}=-\frac{3}{2}\rangle$. They must satisfy the following conditions:
\begin{eqnarray*}
&&|\eta_i\rangle=x_{i}|\hat{B}_2=+\frac{3}{2}\rangle+y_{i}|\hat{B}_2=+\frac{1
}{2}\rangle+z_{i}|\hat{B}_2=-\frac{1}{2}\rangle,\\
&&\rm{at\ least\ two\ of}\ x_{i},\ y_{i},\ z_{i}\not=0\ \rm{for\ a\ given\ i},\  i\ \in \{1,\ 2\},\\
\rm{and} \ && x_{1}^*x_{2}+y_{1}^*y_{2}+z_{1}^*z_{2}=0.
\end{eqnarray*}

We then consider the following product states:
\begin{equation}
\label{prod3by22}
\left.
\begin{array}{lcl}
|\Phi_{15}\rangle=|\hat{A}_2=+\frac{3}{2}\rangle\otimes|\eta_1\rangle,\\
|\Phi_{16}\rangle=|\hat{A}_2=+\frac{3}{2}\rangle\otimes|\eta_2\rangle,\\
|\Phi_{17}\rangle=|\hat{A}_2=+\frac{1}{2}\rangle\otimes|\eta_1\rangle,\\
|\Phi_{18}\rangle=|\hat{A}_2=+\frac{1}{2}\rangle\otimes|\eta_2\rangle,\\
|\Phi_{19}\rangle=|\hat{A}_2=-\frac{1}{2}\rangle\otimes|\eta_1\rangle,\\
|\Phi_{20}\rangle=|\hat{A}_2=-\frac{1}{2}\rangle\otimes|\eta_2\rangle.
\end{array}
\right \}
\end{equation}

For every choice of the four states $|\chi_1\rangle$, $|\chi_2\rangle$, $|\eta_1\rangle$ and $|\eta_2\rangle$, Eq.(\ref{prod3by21}) and Eq.(\ref{prod3by22}) give rise to a set of twelve product states. So, by varying $|\chi_1\rangle$, $|\chi_2\rangle$, $|\eta_1\rangle$ and $|\eta_2\rangle$, we can have infinitely many classes of such twelve product states. However, one can show that only eight of these infinitely many product states in $S^{\independent}$ are linearly independent. In particular, the eight states $\{|\Phi_i\rangle, i:9 \rm{\ to\ }16 \}$, that appeared in Eq.(\ref{prod3by21}) and Eq.(\ref{prod3by22}), are linearly independent. Moreover, one must also note that all the kets $\{|\Phi_i\rangle, i=9 \rm{\ to\ }20 \}$ are orthogonal not only to the subspace $S$, but also to the state $|\Phi_8\rangle$.\\

\subsection{The Hardy subspace}
The most general state $|\Psi\rangle$ satisfying the conditions in Eq.(\ref{prob3by2}), has to be of the form $|\Psi\rangle=v_{0}|\psi\rangle+\displaystyle\sum_{i=9}^{16}v_{i}|\Phi_{i}\rangle$, with $\displaystyle\sum_{i,\ j=9}^{16}v_{i}v_{j}^*\langle\Phi_{j}|\Phi_{i}\rangle+|v_0|^2=1,\ v_0\not=0$, where $|\psi\rangle$ spans the one-dimensional subspace $(S\oplus S')^{\independent}$ of the Hardy subspace $S^{\independent}$. Using Gram-Schmidt orthonormalization method, one can get the unique form of $|\psi\rangle$ as:
\begin{eqnarray}
&&|\psi\rangle=\frac{|\Phi_8\rangle-\displaystyle\sum_{i=1}^7\langle\Phi'_i|\Phi_8\rangle|\Phi'_i\rangle}{\sqrt{1-\displaystyle\sum_{i=1}^7|\langle\Phi'_i|\Phi_8\rangle|^2}},
\label{psimax3by2}
\end{eqnarray}
where $|\Phi'_i\rangle$'s are given by:
\begin{equation*}
\left.
\begin{array}{lcl}
|\Phi'_1\rangle=|\Phi_1\rangle,\\
|\Phi'_2\rangle=|\Phi_2\rangle,\\
|\Phi'_3\rangle=|\Phi_3\rangle,\\
|\Phi'_4\rangle=|\Phi_4\rangle.\\
\end{array}
\right\}
\end{equation*}
\begin{equation*}
\left.
\begin{array}{lcl}
|\Phi'_5\rangle=|\Phi''_5\rangle,\\
|\Phi'_6\rangle=\frac{|\Phi''_6\rangle-\langle\Phi'_5|\Phi''_6\rangle|\Phi'_5\rangle}{\sqrt{1-(|\langle\Phi'_5|\Phi''_6\rangle|^2)}},\\
|\Phi'_7\rangle=\frac{|\Phi''_7\rangle-\langle\Phi'_5|\Phi''_7\rangle|\Phi'_5\rangle-\langle\Phi'_6|\Phi''_7\rangle|\Phi'_6\rangle}{\sqrt{1-(|\langle\Phi'_5|\Phi''_7\rangle|^2+|\langle\Phi'_6|\Phi''_7\rangle|^2)}}.\\
\end{array}
\right\}
\end{equation*}
\begin{equation*}
\left.
\begin{array}{lcl}
\rm{with\ }|\Phi''_5\rangle=\frac{|\Phi_5\rangle-\langle\Phi'_2|\Phi_5\rangle|\Phi'_2\rangle-\langle\Phi'_3|\Phi_5\rangle|\Phi'_3\rangle-\langle\Phi'_4|\Phi_5\rangle|\Phi'_4\rangle}{\sqrt{1-(|\langle\Phi'_2|\Phi_5\rangle|^2+|\langle\Phi'_3|\Phi_5\rangle|^2+|\langle\Phi'_4|\Phi_5\rangle|^2)}},\\
|\Phi''_6\rangle=\frac{|\Phi_6\rangle-\langle\Phi'_2|\Phi_6\rangle|\Phi'_2\rangle-\langle\Phi'_3|\Phi_6\rangle|\Phi'_3\rangle-\langle\Phi'_4|\Phi_6\rangle|\Phi'_4\rangle}{\sqrt{1-(|\langle\Phi'_2|\Phi_6\rangle|^2+|\langle\Phi'_3|\Phi_6\rangle|^2+|\langle\Phi'_4|\Phi_6\rangle|^2)}},\\
|\Phi''_7\rangle=\frac{|\Phi_7\rangle-\langle\Phi'_2|\Phi_7\rangle|\Phi'_2\rangle-\langle\Phi'_3|\Phi_7\rangle|\Phi'_3\rangle-\langle\Phi'_4|\Phi_7\rangle|\Phi'_4\rangle}{\sqrt{1-(|\langle\Phi'_2|\Phi_7\rangle|^2+|\langle\Phi'_3|\Phi_7\rangle|^2+|\langle\Phi'_4|\Phi_7\rangle|^2)}}.
\end{array}
\right\}
\end{equation*}

\subsection{Functional form of the amount of nonlocality}

The amount of nonlocality $q$ for the maximally non-local Hardy state $|\Psi\rangle=|\psi\rangle$ (like in section \ref{func1}), is given by:
\begin{eqnarray}
q&=&|\langle\psi|\Phi_8\rangle|^2\nonumber\\
&=&1- \displaystyle\sum_{i=1}^7|w_{i}|^2,
\label{gen3by27}
\end{eqnarray}
where $w_i=\langle \Phi'_i|\Phi_8\rangle$ for $i \in \{1,\ 2,\ 3,\ 4,\ 5,\ 6,\ 7\}$.

Now, one can find out the values of $|w_{i}|^2$ in terms of the coefficients $a_{jk}$'s and $b_{jk}$'s that appeared in Eq.(\ref{eigenvec3by2}). $q$ then takes the form:
\begin{equation}
\label{q3by2}
\left.
\begin{array}{lcl}
q=(1-(|a_{24}|^2+|a_{34}|^2+|a_{44}|^2))-\frac{|b_{24}|^2(1-(|a_{24}|^2+|a_{34}|^2+|a_{44}|^2))^2}{(1-|b_{24}|^2(|a_{24}|^2+|a_{34}|^2+|a_{44}|^2))}\\
-\frac{|b_{34}|^2(1-(|a_{24}|^2+|a_{34}|^2+|a_{44}|^2))^2}{(1-|b_{24}|^2(|a_{24}|^2+|a_{34}|^2+|a_{44}|^2))(1-(|b_{24}|^2+|b_{34}|^2)(|a_{24}|^2+|a_{34}|^2+|a_{44}|^2))}\\
-\frac{|b_{44}|^2(1-(|a_{24}|^2+|a_{34}|^2+|a_{44}|^2))^2}{(1-(|b_{24}|^2+|b_{34}|^2)(|a_{24}|^2+|a_{34}|^2+|a_{44}|^2))}\frac{1}{(1-((|b_{24}|^2+|b_{34}|^2+|b_{44}|^2)(|a_{24}|^2+|a_{34}|^2+|a_{44}|^2)))}\\
-|a_{14}b_{14}|^2,
\end{array}
\right\}
\end{equation}

Upon substitution of the coefficients in terms of $\theta_1$, $\theta_2$, $\phi_1$ and $\phi_2$, $q$ takes the form:
\begin{equation}
\label{q3by2'}
\left.
\begin{array}{lcl}
q=-\{8\cos^2\frac{\theta_1}{2}\cos^2\frac{\theta_2}{2}(15-8\cos\theta_1+\cos2\theta_1)(15-8\cos\theta_2+\cos2\theta_2)\sin^6\frac{\theta_1}{2}\sin^6\frac{\theta_2}{2}\}\\
/ \{165\cos\theta_1-66\cos2\theta_1+11\cos3\theta_1+30\cos^2\frac{\theta_1}{2}(15-8\cos\theta_1+\cos2\theta_1)\cos2\theta_2\\
-12\cos^2\frac{\theta_1}{2}(15-8\cos\theta_1+\cos2\theta_1)\cos2\theta_2\\
+2\cos^2\frac{\theta_1}{2}(15-8\cos\theta_1+\cos2\theta_1)\cos3\theta_2-270\}.
\end{array}
\right\}
\end{equation}

\subsection{Maximizing $q$ through observables}
The partial derivatives of $q$ (given in Eq.(\ref{q3by2'})) with respect to $\theta_1$ and $\theta_2$ are calculated, $\cos\theta_1$ replaced with $a$, and $\cos\theta_2$ replaced with $b$ ($a,\ b\ \in (-1,\ +1)$), and the following simultaneous equation are solved:
\begin{equation*}
\left.
\begin{array}{lcl}
\frac{\partial q}{\partial \theta_1}=0,\ \frac{\partial q}{\partial \theta_2}=0.\\
\rm{i.e.}\\
(-41+a(3+(-3+a)a)(-30+7a(3+(-3+a)a))+147b+3a(3+(-3+a)a)\\
(14+a(3+(-3+a)a))b-3(1+a)^2(7+(-4+a)a)^2b^2+(1+a)^2(7+(-4+a)a)^2b^3)\\
\times\frac{-3(1-a)^(5/2)\sqrt{1+a}(1-b)^3(1+b)(7+(-4+b)b)}{16(-51+11a(3+(-3+a)a)+(1+a)(7+(-4+a)a)(3+2b(3+(-3+b)b)))^2}=0,\\ \\
3(1-a)^3(1+a)(7+(-4+a)a)(1-b)^{5/2}\sqrt{1+b}(-102+11b(3+(-3+b)b))\\
\times\frac{229-(3+2a(3+(-3+a)a))(1+b)^2(7+(-4+b)b)^2-b(3+(-3+b)b)}{2(16(-51+11a(3+(-3+a)a)+(1+a)(7+(-4+a)a)(3+2b(3+(-3+b)b)))^2)}\\
= 0.\\
\end{array}
\right\}
\end{equation*}
The above equations give rise to the following solution: $(\cos\theta_1,\ \cos\theta_2) = 1-2^{2/3}(3-\sqrt{5})^{1/3},\ 1-2^{2/3}(3-\sqrt{5})^{1/3})$.

Thus we see that the optimal value of the symmetric function $q$ in Eq.(\ref{q3by2}) occurs on the plane $\theta_1=\theta_2$. Now taking $\theta_1=\theta_2=\theta$(say) in Eq.(\ref{q3by2}) and maximizing $q$ over $\theta$, one can see that at that value of $\theta \ \in (0,\pi)$ for which $\cos\theta=1-2^{2/3}(3-\sqrt{5})^{1/3}$, $q$ attains its maximum value, which is equal to $\frac{-11+5\sqrt{5}}{2}\approx 0.0901699$, and the corresponding value of $\theta$ is $116.815$ degrees.

\section{Generic forms of $|\psi\rangle$ and $q$}
\label{general}
From the analyses of the systems of two spin-$\frac{1}{2}$, spin-$1$ and spin-$\frac{3}{2}$ particles so far, we highlight a pattern in the forms of the maximally nonlocal state $|\psi\rangle$ and the amount of nonlocality, $q$, exhibited by $|\psi\rangle$, as a function of the observables (i.e. as a function of $\theta_1,\ \theta_2,\ \phi_1,\ \phi_2$).

The state $|\psi\rangle$ of two spin-$j$ particles (where $j \in \frac{1}{2}N$, $N$ being the set of all positive integers), for a given set of observables ($\hat{A_1}$, $\hat{A_2}$, $\hat{B_1}$ and $\hat{B_2}$), which satisfies the Hardy's nonlocaltiy conditions Eq.(\ref{orthoj1})--Eq.(\ref{orthoj4}) maximally, is given by:
\begin{eqnarray}
 |\psi_{j}\rangle=\frac{|\Phi_{4j+2}\rangle-\displaystyle\sum_{i=1}^{4j+1}\langle\Phi'_i|\Phi_{4j+2}\rangle|\Phi'_{i}\rangle}{\sqrt{1-\displaystyle\sum_{i=1}^{4j+1}|\langle\Phi'_i|\Phi_{4j+2}\rangle|}},\ \rm{where}\ j \ \in \frac{1}{2}N
\ (N\ \rm{being\ the\ set\ of\ all\ positive\ integers}),\\
\label{psimax_gen}
\end{eqnarray}
where $|\Phi'_1\rangle=|\hat{A_1}=+j\rangle\otimes|\hat{B_1}=+j\rangle,\ |\Phi'_2\rangle=|\hat{A_1}=+j-1\rangle\otimes|\hat{B_2}=-j\rangle,...,\ |\Phi'_{2j+1}\rangle=|\hat{A_1}=-j\rangle\otimes|\hat{B_2}=-j\rangle$, while $|\Phi'_{2j+2}\rangle,...,\ |\Phi'_{4j+1}\rangle$ are pairwise orthogonal states in the $(4j+1)$-dimensional subspace $S$ (spanned by the linearly independent vectors $|\hat{A_1}=+j\rangle\otimes|\hat{B_1}=+j\rangle,\ |\hat{A_1}=+j-1\rangle\otimes|\hat{B_2}=-j\rangle,...,\ |\hat{A_1}=-j\rangle\otimes|\hat{B_2}=-j\rangle,\ |\hat{A_2}=-j\rangle\otimes|\hat{B_1}=+j-1\rangle,...,\ |\hat{A_2}=-j\rangle\otimes|\hat{B_1}=-j\rangle$) and are orthogonal to all of $|\Phi'_1\rangle,...,\ |\Phi'_{2j+1}\rangle$.

\begin{figure}[ht]
\centering
\subfigure[\ Plots of $q$ as a function of the angles $\theta_1$ and $\theta_2$]{
\includegraphics[height=9cm,width=11cm]{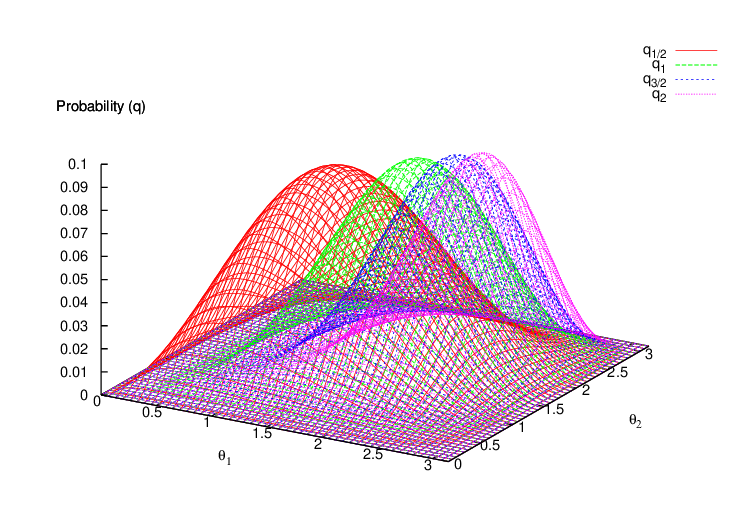}
\label{fig:fig1}
}
\
\subfigure[\ Projection of the plots of $q$ on the $\theta_1=\theta_2$ plane]{
\includegraphics[height=8cm,width=10cm]{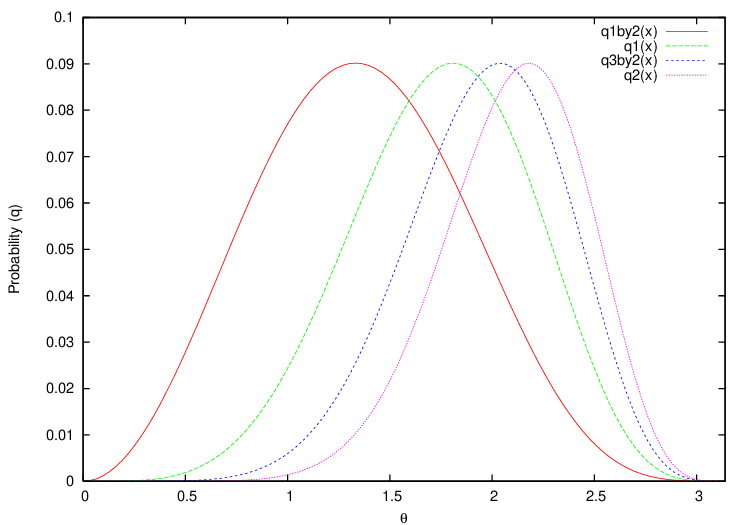}
\label{fig:fig2}
}
\caption{The amounts of nonlocality: $q_{\frac{1}{2}},\ q_1,\ q_{\frac{3}{2}},\ q_2$.}
\end{figure}

The maximum value of $q$ for a given set of observables $\hat{A_1}$, $\hat{A_2}$, $\hat{B_1}$ and $\hat{B_2}$ corresponding to spin-$j \in \frac{1}{2}N$, is given by:
\begin{eqnarray}
&&q_{j}=1-\displaystyle\sum_{i=1}^{4j+1}|\langle\Phi'_i|\Phi_{4j+2}\rangle|^2,
\label{qmax_gen}
\end{eqnarray}
$|\Phi'_i\rangle$'s being described above. The values of $q_{\frac{1}{2}},\ q_1$ and $q_{\frac{3}{2}}$ (given in Eq.(\ref{q1by2}), Eq.(\ref{q1'}) and Eq.(\ref{q3by2'}) respectively) have got a pattern which we assume here to be true for the case $j=2$ also. Appendix D guarantees that, like in the cases for $j=\frac{1}{2},\ 1$ and $\frac{3}{2}$, the maximum value of $q_2$ is also $\frac{-11+5\sqrt{5}}{2} \approx 0.0901699$. Fig.\ref{fig:fig1} shows the plots of $q_j$ values as functions of $\theta_1,\ \theta_2$, for $j=\frac{1}{2},\ 1,\ \frac{3}{2}$ and $2$, while Fig.\ref{fig:fig2} shows the projection of the same plots on the $\theta_1=\theta_2$ plane.

The results described so far for spin values $\frac{1}{2},\ 1,\ \frac{3}{2}$ and $2$ insist us to make the following conjecture.

{\bf Conjecture}: {\it The Hardy's nonlocality conditions given in Eq.(\ref{hardygen}), for any system of two spin-$j$ particles (where $j \in \{\frac{1}{2},\ 1,\ \frac{3}{2}, \ldots\} \equiv (1/2){I\!\!\!N}$, ${I\!\!\!N}$ being the set of all positive integers), with two non-commuting spin-$j$ observables per particle, are satisfied with a maximum value of $q$ over all possible choices of the spin-$j$ observables as $\frac{-11+5\sqrt{5}}{2} \approx 0.0901699$.}

\section{Conclusion}
\label{concl}
Hardy-type nonlocality conditions give rise to contradiction of quantum mechanics with local-realistic description in a straight forward manner without going into statistical inequalities (like Bell's inequalities). However, the Hardy-type nonlocality conditions in general, are weaker than the Bell's inequality violations. For a given pair of non-commuting spin-$j$ observables per site, we have found entangled states $|\psi_j\rangle$ that give rise to maximum nonlocality (in the corresponding Hardy's nonlocality test) for $j=\frac{1}{2}$, $j=1$ and $j=\frac{3}{2}$. We have maximized the nonlocality exhibited by them over all possible choice of the spin-$j$ observables. The maximum nonlocalities have surprisingly turned out to be the same ($\frac{-11+5\sqrt{5}}{2} \approx 0.0901699$) for these three cases. We have figured out a pattern in the form of the maximally nonlocal state and it's maximum amount of nonlocality. The pattern is extended to spin $j=2$ and the result verified (Appendix D). It is conjectured that the maximum amount of nonlocality that can be obtained by the Hardy's nonlocality test on bipartite systems of higher spin values also remains the same. We have also shown that as in the case of a spin-$\frac{1}{2}$ bipartite system, no maximally entangled state of any spin-$j$ bipartite system (where $j \in (1/2){I\!\!\!N}$) responds to the Hardy's nonlocality test. Whether a given non-maximally entangled pure state of two spin-$j$ particles satisfies Hardy's nonlocality conditions can be decided by looking at the SU($2j+1$)$\otimes$SU($2j+1$)-invariants of the states in the Hardy subspace (defined above), which we have described for the cases $j=\frac{1}{2}$, $j=1$.

It is a common belief that nonlocality in quantum mechanics ceases to hold good as we increase the spin value, as that would amount to bringing down the corresponding spin systems nearer to the classical world. However, as in the case of Bell-type inequalities \cite{gisin, peres}, our result shows persistence of nonlocal effect, even if we consider a weaker type of nonlocality, namely, the Hardy-type nonlocality.

The question of finding the states as well as observables for two spin-$j$ particles for the maximal violation of Bell-type inequalities is difficult to answer as there can be several independent Bell-type inequalities in this case (see for example, \cite{lee}), which will be maximally violated by different entangled states, and the amount of maximal violations are also different, in general. Also, these maximal violations will, in general, vary with spin values. This feature will of course depend on the geometry of the state space. We would like to explore in the future, this geometry in the case of the Hardy' nonlocality test for two spin-$j$ particles, specifically for the reason behind getting the same maximum value $\frac{-11+5\sqrt{5}}{2}$. We believe that this feature is typical of {\it all} nonlocality without inequality arguments, not only for the Hardy-type nonlocality argument.

Interestingly, in the line of the conjecture, it might also happen that the maximum value of the nonlocal probability $q$ in the Hardy' nonlocality test for two ($2j+1$)-level particles, will remain one and the same for all non-signalling joint probability distributions (satisfying Hardy-type nonlocality argument given in Eq.(\ref{hardygen})).

\begin{acknowledgments}
K.P.S. thankfully acknowledges Caroline, Meenakshi, Daughty and Purnendu, his colleagues at the 2008-batch summer student program of The Institute of Mathematical Sciences (IMSc), during which the general notion of Hardy's nonlocality argument was discussed. He thanks Sandeep Goyal for useful discussions and help in numerical calculations. K.P.S. also acknowledges BITS, Pilani for deputing him to work at IMSc, and IMSc for it's hospitality. S.G. acknowledges useful discussions with Guruprasad Kar.
\end{acknowledgments}

\section{Appendices}
\begin{center}
\textbf{Appendix A}
\end{center}

Here, we prove that the states $|\Phi_1\rangle$, $|\Phi_2\rangle$, $|\Phi_3\rangle$, $|\Phi_4\rangle$ and $|\Phi_5\rangle$ of Eq.(\ref{ket1}), are linearly independent and span a 5-d subspace $S$, of $C\!\!\!\!I^{\ 3} \otimes C\!\!\!\!I^{\ 3}$.

\textbf{Proof:}
Consider a linear combination of $|\Phi_1\rangle$, $|\Phi_2\rangle$, $|\Phi_3\rangle$, $|\Phi_4\rangle$ and $|\Phi_5\rangle$ equated to zero:
\begin{eqnarray}
&&a|\Phi_1\rangle+b|\Phi_2\rangle+c|\Phi_3\rangle+d|\Phi_4\rangle+e|\Phi_5\rangle=0.\nonumber\\
\Rightarrow
&&a|\hat{A}_1=+1\rangle\otimes|\hat{B}_1=+1\rangle+b|\hat{A}_1=0\rangle\otimes|\hat{B}_2=-1\rangle+c|\hat{A}_1=-1\rangle\otimes|\hat{B}_2=-1\rangle \nonumber\\&&+d|\hat{A}_2=-1\rangle\otimes|\hat{B}_1=0\rangle+e|\hat{A}_2=-1\rangle\otimes|\hat{B}_1=-1\rangle=0.
\end{eqnarray}

Using Eq.(\ref{eigenvec1}) in the above equation, we get:
\begin{eqnarray*}
a(a_{11}|\hat{A}_2=+1\rangle +a_{12}|\hat{A}_2=0\rangle +a_{13}|\hat{A}_2=-1\rangle)\\
\otimes(b_{11}|\hat{B}_2=+1\rangle+b_{12}|\hat{B}_2=0\rangle +b_{13}|\hat{B}_2=-1\rangle)&+&\\
b(a_{21}|\hat{A}_2=+1\rangle +a_{22}|\hat{A}_2=0\rangle +a_{23}|\hat{A}_2=-1\rangle)\otimes|\hat{B}_2=-1\rangle&+&\\
c(a_{31}|\hat{A}_2=+1\rangle +a_{32}|\hat{A}_2=0\rangle +a_{33}|\hat{A}_2=-1\rangle)\otimes|\hat{B}_2=-1\rangle&+&\\
d|\hat{A}_2=-1\rangle\otimes(b_{21}|\hat{B}_2=+1\rangle +b_{22}|\hat{B}_2=0\rangle +b_{23}|\hat{B}_2=-1\rangle)&+&\\
e|\hat{A}_2=-1\rangle\otimes(b_{31}|\hat{B}_2=+1\rangle +b_{32}|\hat{B}_2=0\rangle +b_{33}|\hat{B}_2=-1\rangle)&=&0.
\end{eqnarray*}
$\Rightarrow$
\begin{eqnarray}
|\hat{A}_2=+1\rangle\otimes[aa_{11}b_{11}|\hat{B}_2=+1\rangle+aa_{11}b_{12}|\hat{B}_2=0\rangle+(aa_{11}b_{13}+ba_{21}+ca_{31})|\hat{B}_2=-1\rangle]&+&\nonumber\\|\hat{A}_2=0\rangle\otimes[aa_{12}b_{11}|\hat{B}_2=+1\rangle+aa_{12}b_{12}|\hat{B}_2=0\rangle+(aa_{12}b_{13}+ba_{22}+ca_{32})|\hat{B}_2=-1\rangle]&+&\nonumber\\|\hat{A}_2=-1\rangle\otimes[(aa_{13}b_{11}+db_{21}+eb_{31})|\hat{B}_2=+1\rangle&+&\nonumber\\(aa_{13}b_{12}+db_{22}+eb_{32})|\hat{B}_2=0\rangle+(aa_{13}b_{13}+ba_{23}+ca_{33}+db_{23}+eb_{33})|\hat{B}_2=-1\rangle]&=&0.\nonumber\\
\label{LI}
\end{eqnarray}

Since the eigenvectors $|\hat{A}_2=+1\rangle$, $|\hat{A}_2=0\rangle$ and $|\hat{A}_2=-1\rangle$ are mutually orthogonal, all the three terms on the L.H.S. of Eq.(\ref{LI}) must be separately zero. Thus we get:
\begin{equation}
\label{cond11}
\left.
\begin{array}{lcl}
aa_{11}b_{11}|\hat{B}_2=+1\rangle+aa_{11}b_{12}|\hat{B}_2=0\rangle+(aa_{11}b_{13}+ba_{21}+ca_{31})|\hat{B}_2=-1\rangle=0,\\
aa_{12}b_{11}|\hat{B}_2=+1\rangle+aa_{12}b_{12}|\hat{B}_2=0\rangle+(aa_{12}b_{13}+ba_{22}+ca_{32})|\hat{B}_2=-1\rangle=0,\\
(aa_{13}b_{11}+db_{21}+eb_{31})|\hat{B}_2=+1\rangle+\\
(aa_{13}b_{12}+db_{22}+eb_{32})|\hat{B}_2=0\rangle+(aa_{13}b_{13}+ba_{23}+ca_{33}+db_{23}+eb_{33})|\hat{B}_2=-1\rangle=0.
\end{array}
\right \}
\end{equation}

Since the eigenvectors $|\hat{B}_2=+1\rangle$, $|\hat{B}_2=0\rangle$ and $|\hat{B}_2=-1\rangle$ are also mutually orthogonal, the coefficients of all the three eigenvectors must be separately zero in each of the conditions in Eq.(\ref{cond11}). Since $0<|\langle\hat{A_1}=i|\hat{A_2}=j\rangle|<1$ and $0<|\langle\hat{B_1}=i|\hat{B_2}=j\rangle|<1$, for all $i,j \ \in \{+1,\ 0,\ -1\}$ as per the assumption, $a=0$ and\\
\begin{equation}
\label{summa4}
\left.
\begin{array}{lcl}
ba_{21}+ca_{31}=0,\\
ba_{22}+ca_{32}=0,\\
db_{21}+eb_{31}=0,\\
db_{22}+eb_{32}=0,\\
ba_{23}+ca_{33}+db_{23}+eb_{33}=0.
\end{array}
\right \}
\end{equation}

A nontrivial solution for $b,\ c,\ d,\ e$ exists iff the coefficient matrix $M$ of the above set of simultaneous equations, has a row-rank $\leq 3$, where
\[M=\left(\begin{array}{cccc}
a_{21}& a_{31}& 0& 0\\
a_{22}& a_{32}& 0& 0\\
0& 0& b_{21}& b_{31}\\
0& 0& b_{22}& b_{32}\\
a_{23}& a_{33}& b_{23}& b_{33}\end{array}\right).\]

We now consider the $1^{\rm{st}}$ four rows of $M$. Upon substitution of the $a_{ij}$'s and $b_{ij}$'s in terms of $(\theta_1,\ \phi_1)$ and $(\theta_2,\ \phi_2)$, the reduced matrix gives the following determinant:

\[\left|\begin{array}{cccc}
-\frac{e^{-2i\phi_1} \sin\theta_1}{\sqrt{2}}& \frac{e^{-2i\phi_1} (1-\cos\theta_1)}{2}& 0& 0\\
e^{-i\phi_1}\cos\theta_1& -\frac{e^{-i\phi_1} \sin\theta_1}{\sqrt{2}}& 0& 0\\
0& 0& -\frac{e^{-2i\phi_2} \sin\theta_2}{\sqrt{2}}& \frac{e^{-2i\phi_2} (1-\cos\theta_2)}{2}\\
0& 0& e^{-i\phi_2}\cos\theta_2& -\frac{e^{-i\phi_2} \sin\theta_2}{\sqrt{2}}\end{array}\right|=\frac{e^{-3i(\phi_1+\phi_2)}(2+\cos\theta_1)\sin^2\frac{\theta_1}{2}\sin^2\frac{\theta_2}{2}}{2},\]
which is non-zero for all $\theta_1,\ \theta_2 \ \in (0,\ \pi)$ and $\phi_1,\ \phi_2 \ \in [0,\ 2\pi)$ (including $\theta_1=\frac{\pi}{2}$ and $\theta_2=\frac{\pi}{2}$).

Hence, the rank of $M$ is at least four. Thus, there exists no nontrivial solution for $b,\ c,\ d,\ e$. This proves that the states $|\Phi_1\rangle$, $|\Phi_2\rangle$, $|\Phi_3\rangle$, $|\Phi_4\rangle$ and $|\Phi_5\rangle$ are linearly independent and form a 5-d subspace $S$. \textbf{QED}
\\
\begin{center}
\textbf{Appendix B}
\end{center}

Here, we prove that the states $|\Phi_7\rangle,\ |\Phi_8\rangle,\ |\Phi_9\rangle$ and $|\Phi_{10}\rangle$ of Eq.(\ref{pdt}), span a 3-d subspace $S'$, of $S^{\independent}$.

\textbf{Proof:}
Consider a linear combination of the states $|\Phi_7\rangle,\ |\Phi_8\rangle,\ |\Phi_9\rangle$ and $|\Phi_{10}\rangle$ equated to zero:
\begin{eqnarray}
\alpha|\Phi_7\rangle+\beta|\Phi_8\rangle+\gamma|\Phi_9\rangle+\delta|\Phi_{10}\rangle=0.
\label{gen1}
\end{eqnarray}
\begin{eqnarray}
=>\frac{\alpha}{\sqrt{|b_{11}|^2+|b_{12}|^2}}\left[b_{12}^*|\hat{A}_2=+1\rangle\otimes|\hat{B}_2=+1\rangle-b_{11}^*|\hat{A}_2=+1\rangle\otimes|\hat{B}_2=0\rangle\right]\nonumber\\+\frac{\beta}{\sqrt{|b_{11}|^2+|b_{12}|^2}}\left[b_{12}^*|\hat{A}_2=0\rangle\otimes|\hat{B}_2=+1\rangle-b_{11}^*|\hat{A}_2=0\rangle\otimes|\hat{B}_2=0\rangle\right]\nonumber\\+\frac{\gamma}{\sqrt{|a_{11}|^2+|a_{12}|^2}}\left[a_{12}^*|\hat{A}_2=+1\rangle\otimes|\hat{B}_2=+1\rangle-a_{11}^*|\hat{A}_2=0\rangle\otimes|\hat{B}_2=+1\rangle\right]\nonumber\\+\frac{\delta}{\sqrt{|a_{11}|^2+|a_{12}|^2}}\left[a_{12}^*|\hat{A}_2=+1\rangle\otimes|\hat{B}_2=0\rangle-a_{11}^*|\hat{A}_2=0\rangle\otimes|\hat{B}_2=0\rangle\right]&=&0.
\end{eqnarray}
\begin{eqnarray}
=\rangle \left(\frac{\alpha b_{12}^*}{\sqrt{|b_{11}|^2+|b_{12}|^2}}+\frac{\gamma a_{12}^*}{\sqrt{|a_{11}|^2+|a{12}|^2}}\right)|\hat{A}_2=1\rangle\otimes|\hat{B}_2=1\rangle\nonumber\\+\left(-\frac{\alpha b_{11}^*}{\sqrt{|b_{11}|^2+|b_{12}|^2}}+\frac{\delta a_{12}^*}{\sqrt{|a_{11}|^2+|a_{12}|^2}}\right)|\hat{A}_2=1\rangle\otimes|\hat{B}_2=0\rangle\nonumber\\+\left(\frac{\beta b_{12}^*}{\sqrt{|b_{11}|^2+|b_{12}|^2}}-\frac{\gamma a_{11}^*}{\sqrt{|a_{11}|^2+|a_{12}|^2}}\right)|\hat{A}_2=0\rangle\otimes|\hat{B}_2=1\rangle\nonumber\\+\left(-\frac{\beta b_{11}^*}{\sqrt{|b_{11}|^2+|b_{12}|^2}}-\frac{\delta a_{11}^*}{\sqrt{|a_{11}|^2+|a_{12}|^2}}\right)|\hat{A}_2=0\rangle\otimes|\hat{B}_2=0\rangle&=&0.
\end{eqnarray}

Since the states $|\hat{A}_2=+1\rangle\otimes|\hat{B}_2=+1\rangle$, $|\hat{A}_2=+1\rangle\otimes|\hat{B}_2=0\rangle$, $|\hat{A}_2=0\rangle\otimes|\hat{B}_2=+1\rangle$ and $|\hat{A}_2=0\rangle\otimes|\hat{B}_2=0\rangle$ are pair-wise orthogonal, their coefficients must be separately zero, i.e.,
\begin{equation}
\left.
\begin{array}{lcl}
\frac{\alpha b_{12}^*}{\sqrt{|b_{11}|^2+|b_{12}|^2}}+\frac{\gamma a_{12}^*}{\sqrt{|a_{11}|^2+|a_{12}|^2}}&=&0, \nonumber\\
\frac{\alpha b_{11}^*}{\sqrt{|b_{11}|^2-|b_{12}|^2}}+\frac{\delta a_{12}^*}{\sqrt{|a_{11}|^2+|a_{12}|^2}}&=&0, \nonumber\\
\frac{\beta b_{12}^*}{\sqrt{|b_{11}|^2+|b_{12}|^2}}-\frac{\gamma a_{11}^*}{\sqrt{|a_{11}|^2+|a_{12}|^2}}&=&0, \nonumber\\
\frac{\beta b_{11}^*}{\sqrt{|b_{11}|^2+|b_{12}|^2}}+\frac{\delta a_{11}^*}{\sqrt{|a_{11}|^2+|a_{12}|^2}}&=&0.
\end{array}
\right \}
\end{equation}

The determinant of the coefficient matrix of the above system of linear equations, in the variables $\alpha, \beta, \gamma$ and $\delta$, is:
\[det(M)=\left|\begin{array}{cccc}
b_{12}^*N_1 & 0 & a_{12}^*N_2 & 0 \\
b_{11}^*N_1 & 0 & 0 & -a_{12}^*N_2 \\
0 & b_{12}^*N_1 & -a_{11}^*N_2 & 0 \\
0 & b_{11}^*N_1 & 0 & a_{11}^*N_2 \end{array}\right|,\]
\begin{eqnarray*}
(\rm{where\ } N_1&=&\frac{1}{\sqrt{|b_{11}|^2+|b_{12}|^2}} \ \ ,\ \ N_2=\frac{1}{\sqrt{|a_{11}|^2+|a_{12}|^2}}.),\ \rm{which\ is\ identically\ equal\ to\ 0.}
\end{eqnarray*}

Thus the states $|\Phi_7\rangle$, $|\Phi_8\rangle$, $|\Phi_9\rangle$ and $|\Phi_{10}\rangle$ are not linearly independent. When one of the terms in Eq.(\ref{gen1}) is made equal to zero, say $\delta=0$, the determinant of the coefficient matrix takes the form:
\[detM=\left|\begin{array}{ccc}
b_{12}^*N_1 & 0 & a_{12}^*N_2  \\
b_{11}^*N_1 & 0 & 0  \\
0 & b_{12}^*N_1 & -a_{11}^*N_2 \end{array} \right|=(b_{11}b_{12}a_{12})^*N_1^2N_2,\]
which is non-zero. Hence, it can be concluded that three out of the four states $|\Phi_7\rangle$, $|\Phi_8\rangle$, $|\Phi_9\rangle$ and $|\Phi_{10}\rangle$, are linearly independent and span the 3-d subspace $S'$, of $S^{\independent}$. \textbf{QED}
\\
\begin{center}
\textbf{Appendix C}
\end{center}

The coefficients $a_{ij}$'s and $b_{ij}$'s mentioned in Eq.(\ref{eigenvec3by2}) are as follows:
\begin{equation*}
\left.
\begin{array}{lcl}
a_{11}=\frac{e^{-3 i\phi_1} \cot^3\frac{\theta_1}{2}}{\sqrt{1 + 3 \cot^4\frac{\theta_1}{2}+\cot^6\frac{\theta_1}{2}+3 \cot\theta_1+\csc^2\theta_1}},\\
a_{12}=\frac{\sqrt{3}e^{-2 i\phi_1} \cot^2\frac{\theta_1}{2}}{\sqrt{1 + 3 \cot^4\frac{\theta_1}{2}+\cot^6\frac{\theta_1}{2}+3 \cot\theta_1+\csc^2\theta_1}},\\
a_{13}=\frac{\sqrt{3}e^{-i\phi_1} \cot\frac{\theta_1}{2}}{\sqrt{1 + 3 \cot^4\frac{\theta_1}{2}+\cot^6\frac{\theta_1}{2}+3 \cot\theta_1+\csc^2\theta_1}},\\
a_{14}=\frac{1}{\sqrt{1 + 3 \cot^4\frac{\theta_1}{2}+\cot^6\frac{\theta_1}{2}+3 \cot\theta_1+\csc^2\theta_1}},
\end{array}
\right\}
\end{equation*}
\begin{equation*}
\left.
\begin{array}{lcl}
a_{21}=-\frac{\sqrt{3}e^{-3 i\phi_1}\cot\frac{\theta_1}{2}}{\sqrt{
 3+3\cot^2\frac{\theta_1}{2}+(-3 + \csc^2\frac{\theta_1}{2})^2+(3\cot\theta_1+\csc\theta_1)^2}},\\
a_{22}=\frac{e^{-2 i\phi_1}(-3 + \csc^2\frac{\theta_1}{2})}{\sqrt{
 3+3\cot^2\frac{\theta_1}{2}+(-3 + \csc^2\frac{\theta_1}{2})^2+(3\cot\theta_1+\csc\theta_1)^2}},\\
a_{23}=\frac{e^{-i\phi_1}(3\cot\theta_1+\csc\theta_1)}{\sqrt{
 3+3\cot^2\frac{\theta_1}{2}+(-3 + \csc^2\frac{\theta_1}{2})^2+(3\cot\theta_1+\csc\theta_1)^2}},\\
a_{24}=\frac{\sqrt{3}}{\sqrt{3+3\cot^2\frac{\theta_1}{2}+(-3 + \csc^2\frac{\theta_1}{2})^2+(3\cot\theta_1+\csc\theta_1)^2}},
\end{array}
\right\}
\end{equation*}
\begin{equation*}
\left.
\begin{array}{lcl}
a_{31}=\frac{2 \sqrt{3} e^{-i\phi_1} \tan\frac{\theta_1}{2}}{\sqrt{
(6-\frac{4}{1+\cos\theta_1})^2+4(3+(-1+3\cos\theta_1)^2\csc^2\theta_1+3\tan^2\frac{\theta_1}{2})}},\\
a_{32}=\frac{e^{-2 i\phi_1}(6-\frac{4}{1+\cos\theta_1})}{\sqrt{
(6-\frac{4}{1+\cos\theta_1})^2+4(3+(-1+3\cos\theta_1)^2\csc^2\theta_1+3\tan^2\frac{\theta_1}{2})}},\\
a_{33}=\frac{e^{-i\phi_1}(6\cot\theta_1-2\csc\theta_1)}{\sqrt{
(6-\frac{4}{1+\cos\theta_1})^2+4(3+(-1+3\cos\theta_1)^2\csc^2\theta_1+3\tan^2\frac{\theta_1}{2})}},\\
a_{34}=\frac{2\sqrt{3}}{\sqrt{
(6-\frac{4}{1+\cos\theta_1})^2+4(3+(-1+3\cos\theta_1)^2\csc^2\theta_1+3\tan^2\frac{\theta_1}{2})}},\end{array}
\right\}
\end{equation*}
\begin{equation*}
\left.
\begin{array}{lcl}
a_{41}=-\frac{e^{-3 i\phi_1}\tan^3\frac{\theta_1}{2}}{\sqrt{
 1+3(\cot\theta_1-\csc\theta_1)^2+3\tan^4\frac{\theta_1}{2}+\tan^6\frac{\theta_1}{2}}},\\
a_{42}=\frac{\sqrt{3}e^{-2 i\phi_1}\tan^2\frac{\theta_1}{2}}{\sqrt{
 1+3(\cot\theta_1-\csc\theta_1)^2+3\tan^4\frac{\theta_1}{2}+\tan^6\frac{\theta_1}{2}}},\\
a_{43}=-\frac{\sqrt{3}e^{-i\phi_1}\tan\frac{\theta_1}{2}}{\sqrt{
 1+3(\cot\theta_1-\csc\theta_1)^2+3\tan^4\frac{\theta_1}{2}+\tan^6\frac{\theta_1}{2}}},\\
a_{44}=\frac{1}{\sqrt{
 1+3(\cot\theta_1-\csc\theta_1)^2+3\tan^4\frac{\theta_1}{2}+\tan^6\frac{\theta_1}{2}}},\\
\end{array}
\right\}
\end{equation*}
and similar expressions for the corresponding coefficients $b_{ij}$, with $(\theta_1,\ \phi_1)$ being replaced by $(\theta_2,\ \phi_2)$.
\\
\begin{center}
\textbf{Appendix D}

\textbf{A spin-$2$ bipartite system}
\end{center}

The spin-$2$ observables $\hat{A}_1$ and $\hat{B}_1$ may be written in their most general form, in the eigen-bases of $\hat{A}_2 \equiv \hat{S_z}$ and $\hat{B}_2 \equiv \hat{S_z}$, as:
\[\hat{A}_1=\left( \begin{array}{ccccc}
2\cos\theta_1 & \sin\theta_1 e^{-i\phi_1} & 0 & 0 & 0\\
\sin\theta_1 e^{i\phi_1} & \cos\theta_1 & \frac{\sqrt{3}}{2}\sin\theta_1 e^{-i\phi_1} & 0 & 0\\
0 & 0 & \frac{\sqrt{3}}{2}\sin\theta_1 e^{i\phi_1} & -\cos\theta_1 & \sin\theta_1 e^{-i\phi_1}\\
0 & 0 & 0 & \sin\theta_1 e^{i\phi_1} & -2\cos\theta_1 \end{array} \right) \] \
and a similar expression for $\hat{B}_1$ with $(\theta_1,\ \phi_1)$ replaced by $(\theta_2,\ \phi_2)$.

The expression for $q$ of the maximally nonlocal state of this sytem can be deduced to be of the form:
\begin{equation}
\label{q2'}
\left.
\begin{array}{lcl}
q=\nonumber\\
-\{16\cos^2\frac{\theta_1}{2}\cos^2\frac{\theta_2}{2}(47\cos\theta_1-10(7+\cos2\theta_1)+\cos3\theta_1)\nonumber\\
\times(47\cos\theta_2-10(7+\cos2\theta_2)+\cos3\theta_2)\sin^8\frac{\theta_1}{2}\sin^8\frac{\theta_2}{2}\} \nonumber\\
/ \{-7735-2604\cos2\theta_1+744\cos3\theta_1-93\cos4\theta_1+15680\cos^2\frac{\theta_1}{2}\cos\theta_2\nonumber\\
+4(\cos\theta_1(1302+47\cos^2\frac{\theta_1}{2}(-56\cos\theta_2+28\cos2\theta_2-8\cos3\theta_2+\cos4\theta_2)\nonumber\\
+\cos^2\frac{\theta_1}{2}(-70(28\cos2\theta_2-8\cos3\theta_2+\cos4\theta_2)+(-10\cos2\theta_1+\cos3\theta_1)\nonumber\\
\times(-56\cos\theta_2+28\cos2\theta_2-8\cos3\theta_2+\cos4\theta_2))))\}.
\end{array}
\right\}
\end{equation}

One can now proceed in the direction of section \ref{summa3} and show that the maximum value of $q$ will occur on the plane $\theta_1=\theta_2$. Thus, in order to maximize $q$, we maximize the following function obtained by replacing $\theta_1=\theta_2=\theta$ in the expression for $q$ in Eq.(\ref{q2'}):
\begin{eqnarray}
q&=& \frac{(-70+47\cos\theta-10\cos2\theta+\cos3\theta)^2\sin^8\frac{\theta}{2}\cos^4\frac{\theta}{2}}{8(-221-56\cos\theta+28\cos2\theta-8\cos3\theta+\cos4\theta)}.
\label{q2sym}
\end{eqnarray}

This function maximizes to $\frac{-11+5\sqrt{5}}{2} \approx 0.0901699$, at $\theta_1=\theta_2=\theta=124.9$ degrees.
\end{document}